\documentclass[10pt, twoside]{article} % oneside
\usepackage{robHMM-nospec-header}
\numberwithin{equation}{section}

\newcommand{\ssr}{\rm\scriptscriptstyle}

\newcommand{\iids}{\stackrel{{\tiny\rm i.i.d.}}{\sim}}
\newcommand\e{{\mathbf e}}
\newcommand\E{\mbox{E}}

\newcommand\X{{\mathbf x}}
\newcommand\x{{\mathbf x}}
\newcommand\V{{\mathbf v}}

\newcommand{\Bf}{{\boldsymbol{f}}}

\newcommand{\Bsigma}{{\boldsymbol{\sigma}}}

\newcommand{\Pibold}{{\boldsymbol{\Pi}}}

\title{Robustification of Elliott's on-line EM algorithm for HMMs}
\institute{Fraunhofer ITWM\\
Department of Financial Mathematics\\
Fraunhofer-Platz 1, D-67663 Kaiserslautern\\
\texttt{Christina.Erlwein@itwm.fraunhofer.de}\\
\texttt{Peter.Ruckdeschel@itwm.fraunhofer.de}
}
\author{Christina Erlwein and Peter Ruckdeschel\\
\inst{1}}
\date{\today}

\begin{document}
\maketitle
%-------------------------------------------------------------------------------
\begin{abstract}
In this paper, we establish a robustification of an on-line algorithm for modelling asset prices within a hidden Markov model (HMM). In this HMM framework, parameters of the model are guided by a Markov chain in discrete time, parameters of the asset returns are therefore able to switch between different regimes. The parameters are estimated through an on-line algorithm, which utilizes incoming information from the market and leads to adaptive optimal estimates. We robustify this algorithm step by step against additive outliers appearing in the observed asset prices with the rationale to better handle possible peaks or missings in asset returns. \\
\end{abstract}
%
%-------------------------------------------------------------------------------
\section{Introduction}
Realistic modelling of financial time series from various markets (stocks, commodities, interest rates etc.) in recent years often is achieved through hidden Markov or regime-switching models. One major advantage of regime-switching models is their flexibility to capture switching market conditions or switching behavioural aspects of market participants resulting in a switch in the volatility or mean value.\\
\ \\
Regime-switching models were first applied to issues in financial markets through \cite{H:89}, where he established a Markov switching AR-model to model the GNP of the U.S. His results show promising effects of including possible regime-switches into the characterisation of a financial time series. A lot of further approaches to use regime-switching models for financial time series followed, e.g. switching ARCH or switching GARCH models (see for example \cite{C:94} and \cite{G:96}), amongst many other applications.\\
\ \\
Various algorithms and methods for statistical inference are applied within these model set-ups, including as famous ones as the Baum-Welch algorithm and Viterbi's algorithm for an estimation of the optimal state sequence. HMMs in Finance, both in continuous and in discrete time often utilise a filtering technique which was developed by \cite{E:94}. Adaptive filters are derived for processes of the Markov chain (jump process, occupation time process and auxiliary processes) which are in turn used for recursive optimal parameter estimates of the model parameters. This filter-based Expectation-Maximization (EM) algorithm leads to an on-line estimation of model parameters. Our model set-up is based on Elliott's filtering framework.\\
\ \\
This HMM can be applied to questions, which arise in asset allocation problems. An investor typically has to decide, how much of his wealth shall be invested into which asset or asset class and when to optimally restructure a portfolio. Asset allocation problems were examined in a regime-switching setting by \cite{A:B:02}, where high volatility and high correlation regimes of asset returns were discovered. \cite{G:T:07} presented an asset allocation problem within a regime-switching model and found four different possible underlying market regimes. A paper by \cite{S:H:04} derives optimal trading strategies and filtering techniques in a continuous-time regime-switching model set up. Optimal portfolio choices were also discussed in \cite{E:vdH:97} and \cite{E:H:03} amongst others. Here, Markowitz's famous mean-variance approach (see \cite{Mark:52}) is transferred into an HMM and optimal weights are derived. A similar Markowitz based approach within an HMM was developed in \cite{E:M:D:09}, where optimal trading strategies for portfolio decisions with two asset classes are derived. Trading strategies are developed herein to find optimal portfolio decision for an investment in either growth or value stocks. Elliott's filtering technique is utilised to predict asset returns.\\
\ \\
However, most of the optimal parameter estimation techniques for HMMs in the literature only lead to reasonable results, when the market data set does not contain significant outliers. The handling of outliers is an important issue in many financial models, since market data might be unreliable at times or high peaks in asset returns, which might occur in the market from time to time shall be considered separately and shall not negatively influence the parameter estimation method. In general, higher returns in financial time series might belong to a separate regime within an HMM. This flexibility is already included in the model set-up. However, single outliers, which are not typical for any of the regimes considered, shall be handled with care, a separate regime would not reflect the abnormal data point. In this paper, we will develop a robustification of Elliot's filter-based EM-algorithm. In section \ref{Sec_framework} we will set the HMM framework, which is applied (either in a one- or multi-dimensional setting) to model asset or index returns. The general filtering technique is described in section \ref{Sec_ElliottsAlgorithm}. The asset allocation problem which clarifies the effect outliers can have on the stability of the filters is developed in section \ref{Sec_OutliersInAssetAllocation}. Section \ref{Sec_Robustification} then states the derivation of a robustification for various steps in the filter equations. The robustification of a reference probability measure is derived as well as a robust version of the filter-based EM-algorithm. An application of the robust filters is shown in section \ref{Sec_Implementation} and section \ref{Sec_Conclusion} finishes our work with some conclusions and possible future applications.
\ \\
%-------------------------------------------------------------------------------
%-------------------------------------------------------------------------------
\section{Hidden Markov model framework for asset returns}
\label{Sec_framework}
%-------------------------------------------------------------------------------
For our problem setting we first review a filtering approach for a hidden Markov model in discrete time which was developed by \cite{E:94}. The logarithmic returns of a stock or an index follow the dynamics of the observation process $y_{k},$ which can be interpreted as a discretized version of the Geometric Brownian motion, which is a standard process to model stock returns. The underlying hidden Markov chain $\X_k$ cannot be directly observed. The parameters of the observation process are governed by the Markov chain and are therefore able to switch between regimes over time.\\
\ \\
We
work under a probability space $(\Omega, {\cal F}, P)$ under which $\X_k$ is a homogeneous
Markov chain with finite state space $I=\{1,\ldots,N\}$ in discrete time
$(k=0,1,2...)$. Let the state space of $\X_k$ be associated with the
canonical basis $\{\mathbf{e}_1, \mathbf{e}_2, ...,
\mathbf{e}_N\}\in {\mathbb R}^N$ with
$\mathbf{e}_i=(0,...,0,1,0,...,0)^\top \in {\mathbb R}^N .$
 The initial distribution of $\X_0$
is known and ${\bf\Pi}=(\pi_{ji})$ is the transition probability
matrix with $\pi_{ji}=P(\X_{k+1}=e_j | \X_k=e_i).$ Let ${\cal
F}_k^{\x_0}=\sigma\{\X_0, ..., \X_k\}$ be the $\sigma$-field
generated by $\X_0, ..., \X_k$ and let ${\cal F}_k^{\x}$ be the
complete filtration
generated by ${\cal F}_k^{\x_0}$. Under the real world probability measure $P,$ the Markov chain
${\mathbf x}$ has the dynamics
\begin{equation}
\label{semimartingale}
 {\X}_{k+1}=\Pibold {\X}_k+{\mathbf v}_{k+1}
\end{equation}
where ${\mathbf v}_{k+1}:={\X}_{k+1} - \Pibold {\X}_k$ is a martingale increment (see Theorem in \cite{E:94}).\\
\ \\
The Markov chain $\X_k$ is ``hidden'' in the
log returns $y_{k+1}$ of the stock price $S_k.$ Our observation process is given by
\begin{eqnarray}
\label{StockReturns}
 y_{k+1}=\ln\,\frac{S_{k+1}}{S_k}=f(\X_k)+\sigma(\X_k)w_{k+1}
\end{eqnarray}
where $\X_k$ has finite state space and $w_k$'s constitute a
sequence of i.i.d. random variables independent of $\X.$ The
real-valued process $y$ can be re-written as
\begin{eqnarray}
\label{StockReturns2} y_{k+1}=
\langle
\Bf, \X_k \rangle+\langle\Bsigma, \X_{k}\rangle\,w_{k+1}\,.
\end{eqnarray}
Note that $\mathbf{f}=(f_1,f_2,...,f_N)^\top$ and
$\Bsigma=(\sigma_1, \sigma_2,...,\sigma_N)^\top$ are vectors, furthermore
$f(\X_k)=\langle\mathbf{f},\X_k\rangle$ and
$\sigma(\X_k)=\langle\Bsigma, \X_k\rangle$,
 where $\langle \mathbf{b},\mathbf{c}\rangle$ denotes the Euclidean scalar
product in ${\mathbb R}^N$ of the vectors $\bf{b}$ and $\bf{c}$. We
assume $\sigma_i\neq 0.$ Let ${\cal F}_k^y$ be the filtration
generated by the $\sigma(y_1, y_2,..., y_k)$ and ${\cal F}_k={\cal
F}_k^{\x}\vee{\cal F}_k^y$ is the
global filtration.\\
\ \\
The following theorem (\cite{E:94}) states that
the dynamics of the underlying Markov chain can be described
by martingale differences.
%\begin{Thm}
%Under the real world probability measure $P,$ the Markov chain
%${\mathbf x}$ has the dynamics
%\begin{equation}
%\label{semimartingale}
% {\X}_{k+1}=\Pibold {\X}_k+{\mathbf v}_{k+1}
%\end{equation}
%where
%$\mbox{$\boldmath\Pi$}=(\pi_{ji}),_{j,i=1 \ldots n}$ is the transition probability
%matrix with
%$\pi_{ji}=P({\X}_{k+1}=\e_j | {\X}_k=\e_i)$.  ${\mathbf v}_{k+1}:={\X}_{k+1} - \Pibold {\X}_k$ is a martingale increment.
%\end{Thm}
%\begin{proof}[]
%\begin{eqnarray}
%\E[{\mathbf v}_{k}\mid {\cal F}_{k-1}^{\x}]=\E[\X_{k}-\Pibold {\X}_{k-1}\mid {\X}_{k-1}]\nonumber\\
%&=& \Pibold {\X}_{k-1}-\Pibold {\X}_{k-1}=0\,.\nonumber
%\end{eqnarray}
%
%\end{proof}

\section{Essential Steps in Elliott's Algorithm}
\label{Sec_ElliottsAlgorithm}
%-------------------------------------------------------------------------------
\subsection{Change of Measure}
%-------------------------------------------------------------------------------
A widely used concept in filtering applications, going back to \cite{Z:1969} for stochastic filtering, is a change of probability measure technique. A measure change to a reference measure $\bar P$ is applied here, under which filters for the Markov chain and related processes are derived. Under $\bar P,$ the underlying Markov chain still has the dynamics ${\X}_{k+1}=\Pibold {\X}_k+{\mathbf v}_{k+1}$ but is independent of the observation process and the observations $y_k$ are $\mathcal{N}(0,1)$ i.i.d. random variables.\\
Following the change of measure technique which was outlined in
\cite{E:A:M:95} the adaptive filters for the Markov chain and related processes
are derived under this ``idealised'' measure $\bar{P}.$ Changing back to the real world is done by constructing $P$
from $\bar{P}$ through the
Radon-Nikod\^ym derivative $
\frac{dP}{d\bar{P}}{\bigg |}_{{\cal F}_k}=\Lambda_k. $ To construct
$\Lambda_k$ we define the process $\lambda_l$
\begin{eqnarray}
\label{RNderivativeforGBM} \lambda_l
:=\frac{\phi\Bigl[\sigma(\X_{l-1})^{-1}\bigl(y_l-f(\X_{l-1})\bigr)\Bigr]}{\sigma(\X_{l-1})\phi(y_l)}
\end{eqnarray}
where $\phi (z)$ is the probability density function of a standard
normal random variable $Z$ and set
$\Lambda_k:=\prod_{l=1}^{k} \lambda_l, ~~~k \geq 1, ~~~\Lambda_0=1\,.$
Under $P$ the sequence
of variables $w_1, w_2, \ldots,$ is a sequence of i.i.d. standard
normals,
where we have
$w_{k+1}=\sigma(\X_k)^{-1}\left(y_{k+1}-f(\X_k)\right).$ \\
\ \\
%-------------------------------------------------------------------------------
\subsection{Filtering for general adapted processes}
%-------------------------------------------------------------------------------
The general filtering
techniques and the filter equations which were established by
\cite {E:94} for Markov chains observed in Gaussian
noise are stated in this subsection. This filter-based EM-algorithm is adaptive,
which enables fast calculations and filter updates. Our robustification partly keeps this adaptive structure of the algorithm, although the recursivity cannot be kept completely.\\
In general, filters for four types of processes related to the Markov
chain, namely the state space process, the jump process, the
occupation time process and auxiliary processes including terms of
the observation process are derived. Information on these processes can be filtered out from our observation process and can in turn be used to find optimal parameter estimates.\\
\ \\
To determine the expectation of any ${\cal F}-$adapted stochastic
process $H$ given the filtration ${\cal F}^y_k,$ consider the
reference probability measure $\bar{P}$ defined as
$P(A)=\int_A\, {\Lambda}\,d\overline P\,.$
From Bayes' theorem a filter for any adapted process
$H$ is given by
%\begin{eqnarray*}
$\E\left[H_k \mid {\cal F}_k^y\right]=%\frac
{\overline E\bigl[H_k
{\Lambda}_k\mid {\cal F}_k^y \bigr]}\,\Big/\, {\overline E\bigl[
{\Lambda}_k\mid {\cal F}_k^y \bigr]}~~.$
%\end{eqnarray*}
We define $ \eta_k(H_k):=\overline \E\bigl[H_k {\Lambda}_k\mid
{\cal F}_k^y \bigr] ,$ so that $ \E\bigl[H_k\mid {\cal
F}_k^y\bigr]=%\frac
{\eta_k(H_k)}\,/\,{\eta_k(1)}. $ A recursive
relationship between $\eta_k(H_k)$ and $\eta_{k-1}(H_{k-1})$ has to
be found, where $\eta_0(H_0)=E[H_0].$ However,
a recursive formula for the term $\eta_{k-1}(H_{k-1}\X_{k-1})$ is found.
%Note that $H_k$ is scalar whilst $\eta_{k-1}(H_{k-1}\X_{k-1})$ is a vector.
To relate $\eta_k(H_k)$ and $\eta_k(H_{k}\X_{k})$ we note
that with $\langle \mathbf{1}, \X_k\rangle=1$
\begin{eqnarray}
\label{vectorscalar}
\langle \mathbf{1}, \eta_k(H_k\X_k)
\rangle=\eta_k(H_k\langle\mathbf{1}, \X_k \rangle) =\eta_k(H_k).
\end{eqnarray}
Therefore
\begin{eqnarray} \E\bigl[H_k\mid{\cal
F}_k^y\bigr]&=&\frac{\langle\mathbf{1}, \eta_k(H_k\X_k)
\rangle}{\langle \mathbf{1}, \eta_k(\X_k)\rangle}.
\end{eqnarray}

A general recursive filter for adapted processes was derived by \cite{E:94}.
Suppose $H_l$ is a scalar ${\cal F}=\sigma((\X_t,Y_t)_t)-$adapted process, $H_0$ is
${\cal F}_0^\X$ measurable and
$H_l=H_{l-1}+a_l+\langle b_l,\V_l \rangle+g_lf(y_l),$
where $a,$ $b$ and $g$ are ${\cal F}$-predictable, $f$ is a scalar-valued function and
$\V_l=\X_{l}-\Pi \X_{l-1}.$ A recursive relation for $\eta_k(H_k
\X_k)$ is given by
\begin{eqnarray}
\label{filter} \eta_k(H_k \X_k)&=&
\sum_{i=1}^N\Gamma^i(y_{k})\bigl[\langle \e_i,
\eta_{k-1}(H_{k-1}\X_{k-1}) \rangle\Pi \e_i\nonumber\\
& &~~+\langle \e_i, \eta_{k-1}(a_k\X_{k-1}) \rangle\Pi
\e_i\nonumber\\
& &~~+(\mbox{\textrm{diag}}(\Pi \e_i)-(\Pi \e_i)(\Pi
\e_i)')\eta_{k-1}(b_k\langle \e_i, \X_{k-1}\rangle)\nonumber\\
& &~~+\eta_{k-1}(g_k\langle \e_i, \X_{k-1}\rangle)f(y_k)\Pi \e_i
\bigr]
\end{eqnarray}
Here, for any column vectors $\bf z$ and $\bf y,$ $\bf z \bf
y'$ denotes the rank-one (if $\bf z\not=0$ and $\bf y\not=0$) matrix
$\bf z \bf y^{\top}.$ The term $\Gamma^i(y_{k})$
denotes the component-wise Radon-Nikod\^ym derivative
$\lambda_k^i$:\\
$$\Gamma^i(y_{k})=\phi\bigl(\frac{y_k-f_i}{\sigma_i}\bigr)/ \sigma_i\phi(y_k)$$

%\textbf{General Adapted Processes II }
Now, filters for
the state of the Markov chain as well as for three related
processes: the jump process, the occupation time process and
auxiliary processes of the Markov chain are derived. These processes can be
characterised as special cases of the general process
$H_l.$\\
\ \\
The estimator for the state $\X_k$ is
derived from $\eta_k(H_k\X_k)$ by setting $H_k=H_0=1,$ $a_k=0,$
$b_k=0$ and $g_k=0.$ This implies that
\begin{eqnarray}
\eta_k(\X_{k})=\sum_{i=1}^N\Gamma^i(y_k)\langle \e_i,
\eta_{k-1}(\X_{k-1})\rangle\Pi \e_i~~.
\end{eqnarray}
 The first related process is
the number of jumps of the Markov chain $\X_k$ from state $\e_r$ to
state $\e_s$ in time $k,$
%\begin{eqnarray}
$J_k^{(sr)}=\sum_{l=1}^k \langle \X_{l-1}, \e_r\rangle\langle
\X_l,\e_s\rangle$.
%&=&J_{k-1}^{(sr)}+\langle \X_{k-1}, \e_r\rangle\pi_{sr}+\langle
%\X_{k-1},\e_r\rangle\langle \V_k,\e_s\rangle~~.
%\end{eqnarray}
Setting $H_k=J_k^{(sr)}$,$H_0=0,$ $a_k=\langle \X_{k-1}, \e_r\rangle
\pi_{sr}$,
$b_k=\langle \X_{k-1}, \e_r\rangle \e_s^{'}$ and $g_k=0$ in equation ($\ref{filter}$) we get\\
\begin{eqnarray}
\label{processJ} \eta_k(J_k^{sr}\X_k)
%&=&\sum_{i=1}^N\Gamma^i(y_k)\overline{E}\bigl[\Lambda_{k-1}\langle %%@
%\X_{k-1},\e_i\rangle\{J_{k-1}^{sr}\Pi \e_i+\langle X_{k-1},\e_r\rangle\pi_{sr}\Pi \e_i\nonumber\\
%& &+\langle \X_{k-1},\e_r\rangle \e_s'(\mbox{diag}\Pi \e_i - (\Pi \e_i)(\Pi \e_i)')\}\bigr]\nonumber\\
&=&\sum_{i=1}^N\Gamma^i(y_k)\langle
\eta_{k-1}(J_{k-1}^{sr}\X_{k-1}),\e_i\rangle\Pi \e_i\nonumber\\
& &+\Gamma^r(y_k)\eta_{k-1}(\langle
\X_{k-1},\e_r\rangle)\pi_{sr}\e_s~~.
\end{eqnarray}
\ \\
The second process $O_k^{(r)}$ denotes the occupation time of the
Markov process $\X$, which is the length of time $\X$ spent in state
$r$ up to time $k.$ Here,
%\begin{eqnarray}
$O_k^r=\sum_{l=1}^k\langle \X_{l-1}, \e_r \rangle=O_{k-1}^r+\langle
\X_{k-1}, \e_r\rangle\,.$
%\end{eqnarray}
We set $H_k=O_k^r$, $H_0=0,$ $a_k=\langle \X_{k-1}, \e_r \rangle$,
$b_k=0$ and $g_k=0$ in equation ($\ref{filter}$) to obtain
\begin{eqnarray}
\label{processO} \eta_k(O_{k}^r\X_k)
%&=&\sum_{i=1}^N\Gamma^i(y_k)\{\langle \eta_{k-1}(O_{k-1}^r\X_{k-1}),\e_i\rangle\Pi \e_i\nonumber\\
%& &+\eta_{k-1}(\langle \X_{k-1},\e_r\rangle\langle
%\X_{k-1},\e_i\rangle)\Pi \e_i\}\nonumber\\
&=& \sum_{i=1}^N\Gamma^i(y_k)\langle \eta_{k-1}(O_{k-1}^r\X_{k-1}),\e_i\rangle\Pi \e_i\nonumber\\
& &+\Gamma^r(y_k)\langle \eta_{k-1}(\X_{k-1}),\e_r\rangle\Pi \e_r~~.
\end{eqnarray}
Finally, consider the auxiliary process $T_k^r(g)$, which occur in
the maximum likelihood estimation of model parameters. Specifically,
%\begin{eqnarray}
$T_k^{(r)}(g)=\sum_{l=1}^k\langle \X_{l-1}, \e_r \rangle
g(y_l),$
%&=& T_{k-1}^r(g)+\langle \X_{k-1}, \e_r \rangle g(y_k)
%\end{eqnarray}
where $g$ is a function of the form $g(y)=y$ or $g(y)=y^2.$
%$g(y)=y_{l+1}$, $g(y)=y_{l+1}y_l$ or $g(y)=y_{l+1}^2, 1\leq l\leq k.$
We apply formula (\ref{filter})
%with the substitution $H_k=T_k^r(g),$ $H_0=0,$ $a_k=0$, $b_k=0$ and $g_k=\langle \X_{k-1},\e_r \rangle$
and get
\begin{eqnarray}
\label{processT} \eta_k(T_k^r(g)\X_k)
%&=&\sum_{i=1}^N\Gamma^i(y_k)\{\langle \eta_{k-1}(T_{k-1}^r(g)\X_{k-1}),\e_i\rangle\Pi %%@
%\e_i\nonumber\\
%& &+\eta_{k-1}(\langle \X_{k-1},\e_r\rangle\langle
%\X_{k-1},\e_i\rangle)g(y_k)\Pi \e_i\}\nonumber\\
&=&\sum_{i=1}^N\Gamma^i(y_k)\langle
\eta_{k-1}(T_{k-1}^r(g(y_{k-1}))\X_{k-1}),\e_i\rangle\Pi \e_i\nonumber\\
& &+\Gamma^r(y_k)\langle \eta_{k-1}(\X_{k-1}),\e_r\rangle g(y_k)\Pi
\e_r\,.
\end{eqnarray}
The recursive optimal estimates of $J,$ $O$ and $T$ can be
calculated using equation ($\ref{vectorscalar}$).\\
\subsection{Filter-based EM-algorithm}
The derived adapted filters for processes of the Markov chain can now be utilised to derive optimal parameter estimates through a filter-based EM-algorithm. The set of parameters $\rho$, which determines the regime-switching
model is
\begin{equation}
\label{parameterset} \rho=\{\pi_{ji}, 1\leq i, j\leq N, f_i, \sigma_i,
1\leq i \leq N\}.
\end{equation}
Initial values for the EM algorithm are assumed to be given. Starting from these values updated parameter estimates are derived
which maximise the conditional expectation of the log-likelihoods. The M-step of the algorithm deals with maximizing the following likelihoods:\\
\textbf{M-Step}
\begin{itemize}
\item The likelihood in the global ${\cal F}$-model is given by
$$\log \Lambda_t(\sigma, f;(\X_s,y_s)_{s\leq t})=- \frac{1}{2} \sum_{s=1}^t \Big( \log \langle \sigma, \X_{s-1}\rangle +
\frac{(y_s-\langle  f, \X_{s-1}\rangle)^2}{\langle \sigma, \X_s\rangle} \Big)$$
\item In the ${\cal F}_y$-model, where the Markov chain is not observed, we obtain
\begin{eqnarray}
&&L_t(\sigma, f;(y_s)_{s\leq t}\, = \Ew [\log \Lambda_t(\sigma, f;(\X_s,y_s)_{s\leq t})\,\mid\,{\cal F}_t] = \nonumber\\
&=& - \frac{1}{2} \sum_{k=1}^N \Big(\log \sigma_k \hat O_t^{k}  + (\hat T_t^{k}(y^2) - 2 \hat T_t^{k}(y) f_k + \hat O_t^k  f^2)/\sigma_k^2\Big)
\label{llik}
\end{eqnarray}
%\item for $\hat J_t^{ji}$, $\hat O_t^{i}$, and $\hat T_t^{i}(f)$, obtain MLEs by maximizing $L_t$.
%\item similarly, determine MLE for $\Pibold$
\end{itemize}
The maximum likelihood estimates of the model parameters can be expressed through the adapted filters. Whenever new information is available on the market, the filters are updated and, respectively, updated parameter estimates can be obtained.\\
%Therefore, this hidden Markov model is self-tuning, new information on asset prices can on-line be used for updated parameter estimates.
\ \\
\begin{Thm}[Optimal parameter estimates]
\label{theoremParamEst}
Write $\hat{H}_k=E[H_k|{{\cal
F}^y_k}]$ for any adapted process $H.$
With $\hat J$, $\hat O$ and $\hat T$ denoting the best estimates for
the processes $J$, $O$ and $T$, respectively, the optimal parameter
estimates $\hat{\pi}_{ji}, \hat{f}_i$ and $\hat{\sigma}_i$ are given by
\begin{eqnarray}
\label{pihat}
\hat{\pi}_{ji}&=&\frac{\hat{J}^{ji}_k}{\hat{O}^i_k} = \frac{\eta_k(J^{ji}_k)}{\eta_k(O^i_k)}\\
\label{fhat}
\widehat f_i &=& \frac{\widehat T_k^{(i)}}{\widehat
O_k^{(i)}} =\frac{\eta (T^{(i)}(y))_k}{\eta (O^{(i)})_k}\\
\label{sigmahat}
 \widehat \sigma_i&=&\sqrt{\frac{\widehat T_k^{(i)}(y^2)-2{\widehat f}_i
\widehat T_k^{(i)}(y)+{\widehat f}_i^2 \widehat O_k^{(i)}} {\widehat
O_k^{(i)}}}.
\end{eqnarray}
\end{Thm}
\begin{proof} The derivation of the optimal parameter estimates can be found in \cite{E:A:M:95}.
\end{proof}
%-------------------------------------------------------------------------------
%\subsection{M-Step}
%-------------------------------------------------------------------------------

\textbf{Summary}
The filter-based EM-algorithm runs in batches of $n$ data points ($n$ typically equals to a minimum of ten up to a maximum of fifty) over the given time series. The parameters are updated at the end of each batch.
Elliott's Algorithm comprises the following steps
\begin{enumerate}
\label{ElliotAlgo}
\item[(0)] Find suitable starting values for $\Pibold$ and $ f$, $\sigma.$
\item[(RN)] Determine the RN-derivative for the measure change to $\bar P.$
\item[(E)] Recursively, compute filters $\hat J_t^{ji}$, $\hat O_t^{i}$, and $\hat T_t^{i}(g).$
\item[(M1)] Obtain ML-estimators $\hat  f=(\hat  f_1,\ldots,\hat f_N)$ and $\hat \sigma=(\hat \sigma_1,\ldots,\hat\sigma_N).$
\item[(M2)] Obtain ML-estimators $\Pibold.$
\item[(Rec)] Go to (RN) to compute the next batch.
\end{enumerate}

%
%-------------------------------------------------------------------------------
\section{Outliers in Asset Allocation Problem}
\label{Sec_OutliersInAssetAllocation}
%-------------------------------------------------------------------------------
\subsection{Outliers in General}
%-------------------------------------------------------------------------------
In the following sections we derive a robustification of the Algorithm~\ref{ElliotAlgo}
to stabilize it in the presence of outliers in the observation process.
To this end let us discuss what makes an observation an outlier. First of
all, outliers are exceptional events, occurring \textbf{rarely}, say with
probability $5\%$--$10\%$. Rather than captured by usual randomness,
i.e., by some distributional model, they belong to what \cite{Kn:21} refers
to \textit{uncertainty}: They are \textbf{u}ncontrollable, of
\textbf{u}nknown distribution, \textbf{u}npredictable, their distribution
may change from observation to observation, so they are non-recurrent and do
not form an additional state, so cannot be used to enhance predictive power,
and, what makes their treatment difficult, they often cannot be told with
certainty from ideal observations.\\
\ \\
Still, the majority of the observations in a realistic sample should resemble an
ideal (distributional) setting closely, otherwise the modeling would be questionable.
Here we understand closeness as in a distributional sense, as captured, e.g., by goodness-of-fit
distances like Kolmogorov, total variation or Hellinger distance. More precisely,
ideally, this closeness should be compatible to the usual convergence mode
of the Central Limit Theorem, i.e., with weak topology. In particular,
closeness in moments is incompatible with this idea.\\
\ \\
Topologically speaking, one would most naturally use balls around a certain element,
i.e., the set of all distributions  with a suitable distance no larger than some given radius $\ve>0$
to the distribution assumed in the ideal model.\\
%Unfortunately distances which exactly metricize weak topology,
%like Prokhorov distance, are all but handy, so instead one
%usually passes over to simpler neighborhoods which still are able to capture
%outliers phenomena well.
\ \\
Conceptually, the most tractable neighborhoods are given by the so-called
\textit{Gross Error Model}, defining a neighborhood ${\cal U}$ about
a distribution $F$ as the set of all distributions given by
\begin{equation} \label{GEN}
{\cal U}_c(F,\ve)=\{G\,|\,\exists H\colon G=(1-\ve)F+\ve H\}
\end{equation}
They can also be thought of as the set of all distributions of
(\textit{re}alstic) random variables $X^{\ssr re}$ constructed as
\begin{equation}\label{GEdef}
X^{\ssr re}=(1-U)X^{\ssr id}+ U X^{\ssr di}
\end{equation}
where $X^{\ssr id}$ is a random variable distributed according to
the \textit{id}eal distribution and $U$ is an independent ${\rm Bin}(1,\ve)$
switching variable, which in most cases lets you see $X^{\ssr id}$ but
in some cases replaces it by some contaminating or \textit{di}storting variable $X^{\ssr di}$
which has nothing to do with the original situation.
%-------------------------------------------------------------------------------
\subsection{Time-dependent Context: Exogenous and Endogenous Outliers}
%-------------------------------------------------------------------------------
In our time dependent setup in addition to the i.i.d. situation,
we have to distinguish whether the impact of an outlier is
propagated to subsequent observations or not.
Historically there is a common terminology due to  \cite{F:72}, who distinguishes
{\em innovation outliers} (or IO's) and {\em additive outliers} (or AO's).
Non-propagating AO's are added at random to single observations, while IO's
denote gross errors affecting the innovations. For consistency with literature,
we use the same terms, but  use them in a wider sense:
\textit{IO}'s stand for general endogenous outliers entering the state
layer (or the Markov chain in the present context),
hence with propagated distortion.
As in our Markov chain, the state space is finite, IO's are much
less threatening as they are in general.\\
\ \\
Correspondingly, wide-sense \textit{AO}'s denote general exogenous outliers
which do not propagate, hence also comprise substitutive outliers or
\textit{SO}'s as defined in a simple generalization of \eqref{GEdef}
to the state space context in equations~\eqref{YSO}--\eqref{U-SO}.
\begin{equation}
 Y^{\rm\SSs re} = (1-U)  Y^{\rm\SSs id} + U Y^{\rm\SSs di}, \qquad U\sim {\rm Bin}(1,r) \label{YSO}
\end{equation}
for $U$ independent of $(X,Y^{\rm\SSs id},Y^{\rm\SSs di})$ and some arbitrary distorting
random variable $Y^{\rm\SSs di}$ for which we assume
 \begin{equation}\label{indep2}
 Y^{\rm\SSs di},\; X\quad \mbox{independent}
 \end{equation}
and the law of which is arbitrary, unknown
and uncontrollable. As a first step consider the set $\partial{\cal U}^{\rm\SSs SO}(r)$ defined
as
\begin{equation}
\partial{\cal U}^{\rm\SSs SO}(r)=\Big\{{\cal L}(X,Y^{\rm\SSs re}) \,|\, Y^{\rm\SSs re} \;\mbox{acc. to \eqref{YSO} and \eqref{indep2}}\Big\}
\end{equation}
Because of condition~\eqref{indep2}, in the sequel we refer to the random variables $Y^{\rm\SSs re}$ and $Y^{\rm\SSs di}$ instead of their
respective (marginal) distributions only, while in the common gross error model, reference to the respective distributions would suffice.
  Condition~\eqref{indep2}
 also entails that in general, contrary to the gross error model,
$\Lw(X,Y^{\rm\SSs id})$ is not element of  $\partial{\cal U}^{\rm\SSs SO}(r)$, i.e., not representable itself as some $\Lw(X,Y^{\rm\SSs re})$
in this neighborhood.\\
\ \\
As corresponding (convex) neighborhood we define
\begin{equation} \label{U-SO}
{\cal U}^{\rm\SSs SO}(r)= \bigcup_{0\leq s \le r} \partial{\cal U}^{\rm\SSs SO}(s)
\end{equation}
hence the symbol ``$\partial$'' in $\partial{\cal U}^{\rm\SSs SO}$, as the latter can be interpreted as the corresponding surface of this ball.
Of course, ${\cal U}^{\rm\SSs SO}(r)$ contains $\Lw(X,Y^{\rm\SSs id})$.
In the sequel where clear from the context we drop the superscript ${\rm\Ts SO}$ and the
argument $r$.\\
\ \\
Due to their different nature, as a rule, IO's and AO's require
different policies: As AO's are exogenous, we would
like to damp their effect, while when there are IO's, something has
happened in the system, so  the usual goal will be to  track these
changes as fast as possible.
%-------------------------------------------------------------------------------
\subsection{Evidence for Robustness Issue in Asset Allocation}
\begin{figure}
\begin{center}
\includegraphics[bb=59 212 543 602,height=7cm]{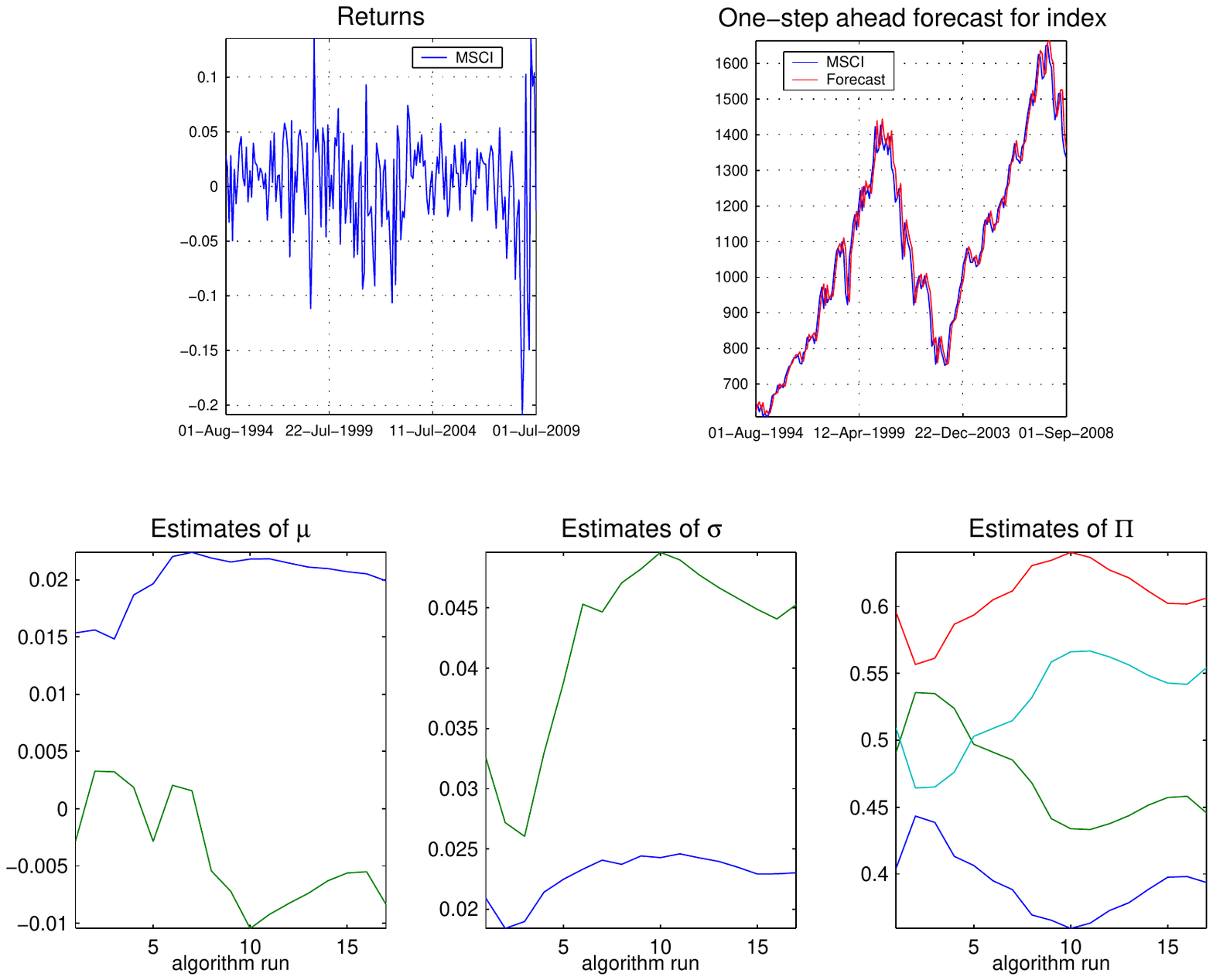}
\caption{Optimal parameter estimates for monthly MSCI returns between 1994 and 2009}
\label{MSCIRet}
\end{center}
\end{figure}

\begin{figure}
\begin{center}
\includegraphics[bb=59 212 543 602,height=7cm]{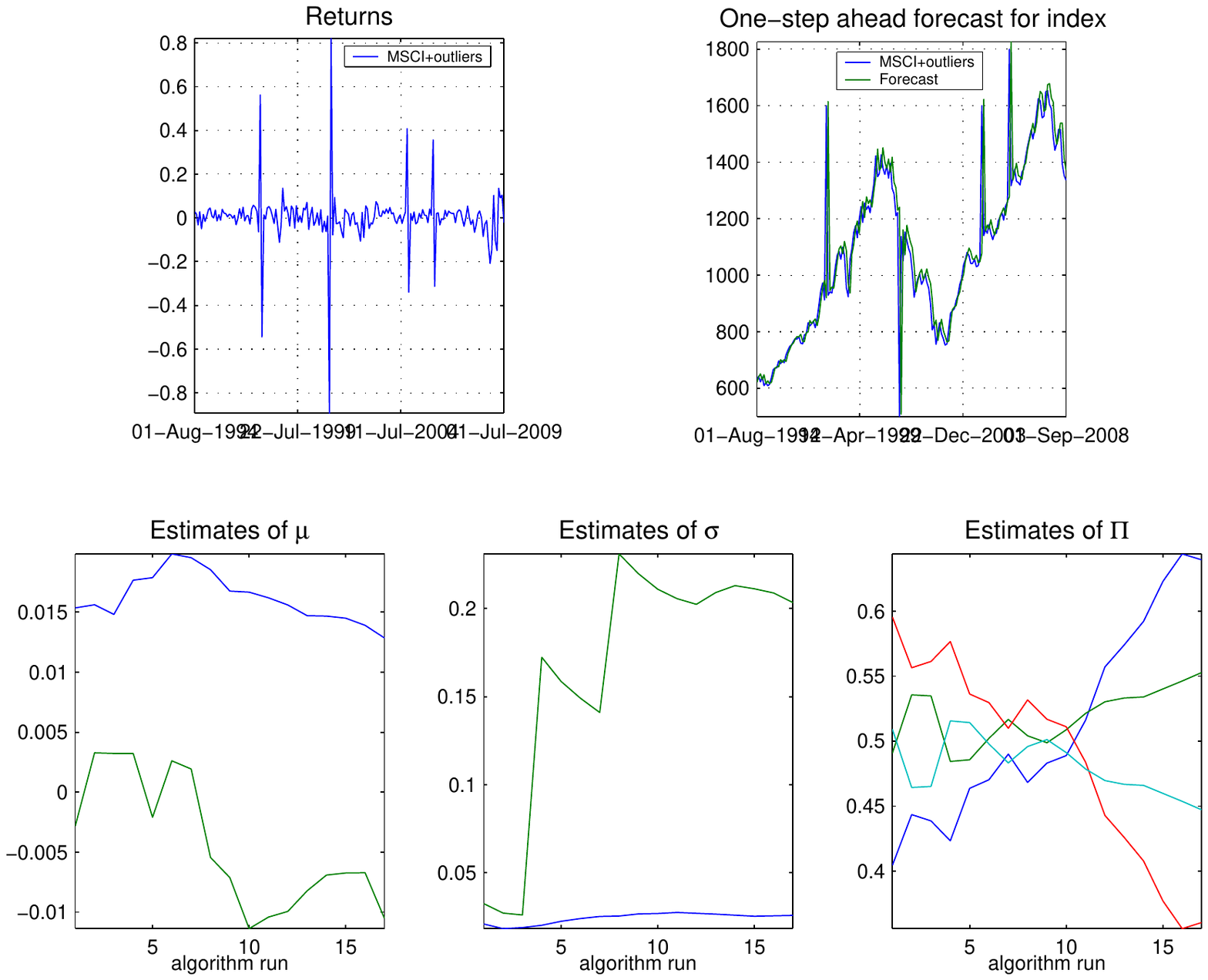}
\caption{Optimal parameter estimates for monthly MSCI returns with planted outliers}
\label{MSCIRetOut1}
\end{center}
\end{figure}

\begin{figure}
\begin{center}
\includegraphics[bb=59 212 543 602,height=7cm]{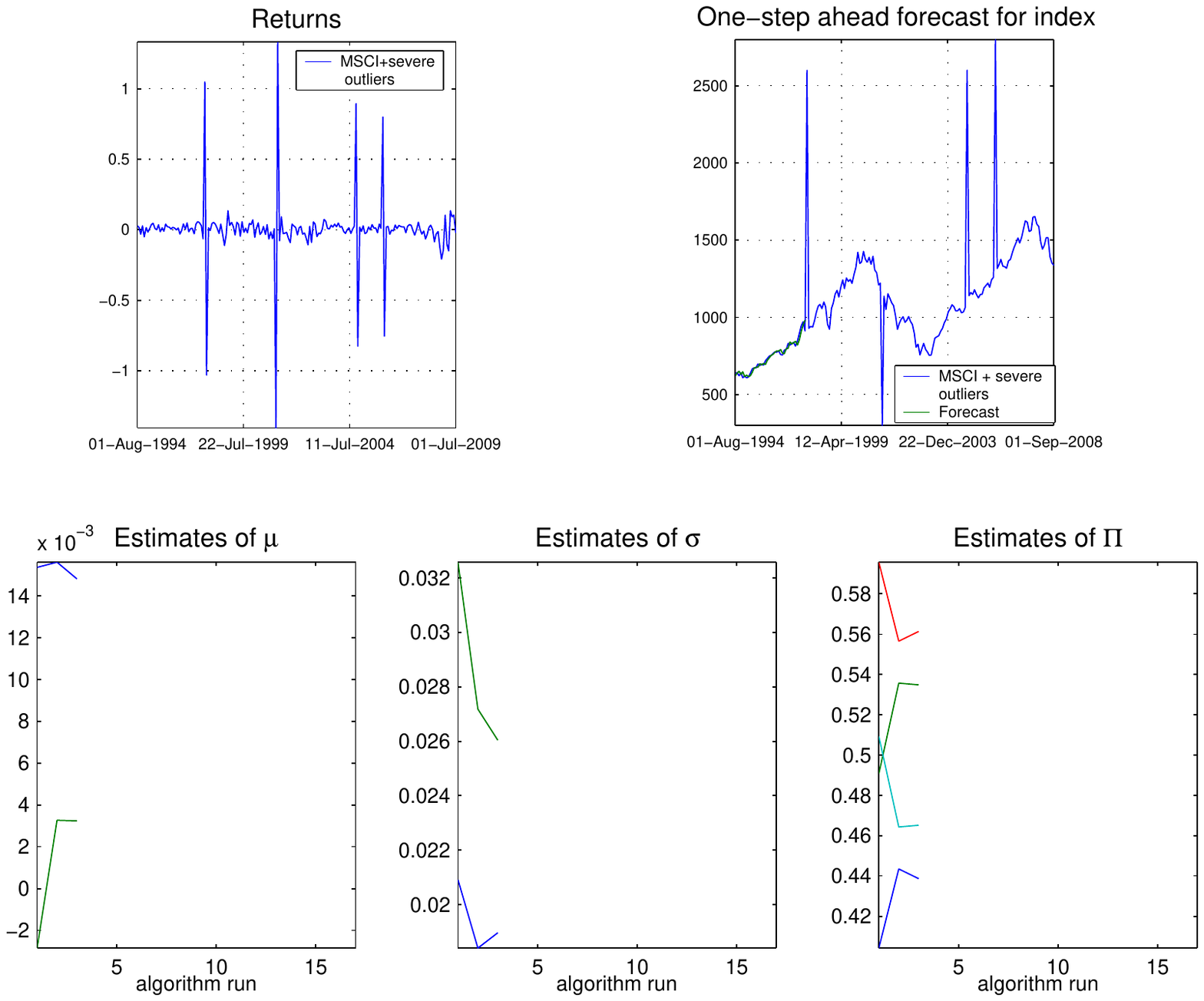}
\caption{Filter-based EM-algorithm for observation sequence with severe outliers}
\label{MSCIRetOut2}
\end{center}
\end{figure}

%-------------------------------------------------------------------------------
In this section we examine the robustness of the filter and parameter estimation technique. The filter technique is implemented and applied to monthly returns of MSCI index between 1994 and 2009. The MSCI World Index is one of the leading indices on the stock markets and a common benchmark for global stocks. The algorithm is implemented with batches of ten data points, therefore the adaptive filters are updated whenever ten new data points are available on the market. The recursive parameter estimates, which utilise this new information, are updated as well, the algorithm is self-tuning. Care has to be taken when choosing the initial values for the algorithm, since the EM-algorithm in its general form converges to a local maximum. In this implementation we choose the initial values with regard to mean and variance of the first ten data points. Figure \ref{MSCIRet} shows the original return series, optimal parameter estimates for the index returns as well as the one-step ahead forecast.\\
\ \\
To highlight the sensitivity of the filter technique towards exogenous outliers we plant unusual high returns within the time series. Considerable SO outliers are included at time steps $t=40,80,130,140.$ The optimal parameter estimation through the filter-based EM-algorithm of this data set with outliers can be seen in Figure \ref{MSCIRetOut1}. The filter still finds optimal parameter estimates, although the estimates are visibly affected by the outliers.\\
\ \\
In a third step, severe outliers are planted into the observation sequence. Data points $t=40,80,130,140$ now show severe SO outliers as can be seen from the first panel in Figure \ref{MSCIRetOut2}. The filters cannot run through any longer, optimal parameter estimates cannot be established in a setting with severe outliers.\\
\ \\
In practice, asset or index return time series can certainly include outliers
from time to time. This might be due to wrong prices in the system, but also
due to very unlikely market turbulence for a short period of time. It has to be noted,
that the type of outliers which we consider in this study does not characterise
an additional state of the Markov chain. In the following, we develop robust
filter equations, which can handle exogenous outliers.
%-------------------------------------------------------------------------------
\section{Robust Statistics}
\label{Sec_Robustification}
%-------------------------------------------------------------------------------
To overcome effects like in Figure~\ref{MSCIRetOut2} we need more
stable variants of the Elliott type filters discussed so far. This
is what robust statistics is concerned with.
Excellent monographs on this topic are e.g., \cite{Hu:81}, \cite{Ha:Ro..:86},
\cite{Ried:94}, \cite{M:M:Y:06}.
This section provides necessary concepts and results from
robust statistics needed to obtain the optimally-robust estimators
used in this article.

%-------------------------------------------------------------------------------
\subsection{Concepts of Robust Statistics}
%-------------------------------------------------------------------------------

The central mathematical concepts of continuity, differentiability, or
closeness to singularities may in fact serve to operationalize stability
quite well already. To make these available in our context, it helps to consider
a statistical procedure, i.e.; an estimator, a predictor, a filter, or a test
as a function of the underlying distribution. In a parametric context, this
amounts to considering functionals $T$ mapping
%(a subset of all)
distributions
%(at least all model distributions $F_\theta$)
to the parameter set $\Theta$.
An estimator will then simply be $T$ applied to the  empirical distribution $\hat F_n$.
For filtering or prediction, the range of such a functional will rather
be the state space but otherwise the arguments run in parallel.

For a notion of continuity, we have to specify a topology, and as in case
of outliers, we use topologies essentially compatible with the
weak topology.
With these neighborhoods, we now may easily translate the notions
of continuity, differentiability and closest singularity to this context:
(Equi-)continuity is then called
\textit{qualitative robustness} \cite[Sec.~2.2~Def.~3]{Ha:Ro..:86},
a differentiable functional with a bounded derivative is called
\textit{local robust},
and its derivative is called \textit{influence function}
(IF)\footnote{In mathematical rigor, the IF, when it exists, is the G\^ateaux
derivative of functional $T$ into  the direction of the
tangent $\delta_x-F$. For certain properties, this notion is in fact too weak,
and one has to require stronger notions like Hadamard or Fr\'echet differentiability;
for details, see \cite{Fe:83} or \cite[Ch.~1]{Ried:94}.}, compare
\cite[Sec.~2.1~Def.~1]{Ha:Ro..:86}.
The IF reflects the infinitesimal influence
of a single observation on the estimator.
Under additional assumptions, many of the asymptotic properties of
an estimator are expressions in the IF $\psi$. E.g., the asymptotic variance of
the estimator in the ideal model
is the second moment of $\psi$. Infinitesimally, i.e., for $\ve\to0$,
the maximal bias on ${\cal U}$  is just $\sup|\psi|$, where $|\cdot|$ denotes
Euclidean norm. $\sup|\psi|$ is then also called \textit{gross error sensitivity} (GES),
\cite[(2.1.13)]{Ha:Ro..:86}. Seeking robust optimality hence amounts to finding
optimal IFs.

To grasp the maximal bias of a functional $T$ on a neighborhood ${\cal U}={\cal U}(F;\ve)$ of radius $\ve$,
one considers the \textit{max-bias curve} $\ve \mapsto \sup_{G\in{\cal U}(F;\ve)}|T(G)-T(F)|$.
The singularity of this curve closest to 0 (i.e., the ideal situation of no outliers) captures
its behavior under massive deviations, or its global robustness.
In robust statistics, this is called \textit{breakdown point}---the
maximal radius $\ve$ the estimator can cope with without producing an arbitrary large bias, see
\cite[Sec.~2.2~Def.'s~1,2]{Ha:Ro..:86} for formal definitions.

Usually, the classically optimal estimators (MLE in many circumstances)
are non-robust, both locally and globally. Robust estimators on the other
hand pay a certain price for this stability as expressed by
an \textit{asymptotic
relative efficiency} (ARE) strictly lower than $1$ in the ideal model,
where ARE is the ratio of the two
%(traces of the)
asymptotic (co)variances of the classically optimal estimator and its
robust alternative.

To rank various robust procedures among themselves, other
quality criteria are needed, though,
summarizing the behavior of the procedure on a whole neighborhood,
as in \eqref{GEN}. A natural candidate for such a criterion is
maximal MSE (maxMSE) on some neighborhood
${\cal U}$ around the ideal model and, for estimation context,
maximal bias (maxBias) on the respective neighborhood, or,
referring to famous \cite[Lemma~5]{Ha:68}, trace of the ideal variance
subject to a bias bound on this neighborhood. In estimation context,
the respective solution are usually called OMSE (\textit{O}ptimal
\textit{MSE} estimator), MBRE (\textit{M}ost \textit{B}ias
\textit{R}obust \textit{E}stimator), and OBRE (\textit{O}ptimally \textit{B}ias  \textit{R}obust
\textit{E}stimator)\footnote{For the terms OBRE and MBRE, see
\cite{Ha:Ro..:86}, while for OMSE see \cite{R:H:10}.}.

In our context we encounter two different situations where
we want to apply robust ideas: (recursive) filtering in the
(E)-step and estimation in the (M)-step.
While in the former situation we only add a single new observation,
which precludes asymptotic arguments, in the (M)-step,
the preceding application of Girsanov's theorem turns our situation into an
 i.i.d. setup,
where each observation becomes (uniformly) asymptotically negligible and
asymptotics apply in a standard form.

%For its definition, OMSE requires the radius $r$ of the
%neighborhood to be known, which is almost never the case in practice.
%When the radius is unknown (or only partially known), the
%\textit{radius-minimax estimator} ({RMXE}) introduced in \cite{Ried:07}
%provides a certain optimality even then:
%For any (arbitrarily fixed) radius $s$ and fixed procedure ${\rm OMSE}_s$
%(optimal for radius $s$), we vary the true radius $r$ and determine the maximal
%efficiency loss in terms of relative maxMSE in relation to the best procedure
%knowing the true radius $r$ (i.e., ${\rm OMSE}_r$) and then,
%in an outer loop minimize this maximal efficiency, varying $s$.
%This gives a least favorable radius $s=r_{lf}$ for the neighborhood.
%The estimator optimal on the neighborhood of this radius $r_{lf}$, i.e.,
%${\rm OMSE}_{r_{lf}}$, is called
%and is recommended.
%%-------------------------------------------------------------------------------
\subsection{Our Robustification of the HMM: General Strategy}
%%-------------------------------------------------------------------------------
%
As a robustification of the whole estimation process in this only
partially observed model would (a) lead to computationally intractable
terms and (b) would drop the key feature of recursivity, we instead
propose to robustify each of the steps in Elliott's algorithm separately.
Doing so, the whole procedure will be robust, but in general will loose
robust optimality, i.e.;  contrary to multiparameter maximum likelihood,
the Bellmann principle does not hold for optimal robust multi-step procedures
simply because optimal clipping in several steps is not the same as
joint optimal clipping of all steps.
Table \ref{TabAlgo} lists all essential steps in Elliot's algorithm
related to our proposed robustification approach.\\
\ \\
\begin{table}[ht]
\center
\begin{tabular}{l|l}
  \textbf{Classical setting} & \textbf{Robust version}\\
  \hline
  \multicolumn{2}{l}{}\\
  \multicolumn{2}{l}{\textbf{Initialization:} Find suitable starting values for $\Pibold,$ $ f$, $\sigma$ and $\X_0.$}\\
  \multicolumn{2}{l}{}\\
  \hline
  Build $N$ clusters on first batches     & Build $N+1$ clusters on first batches, \\
                                        & distribute points in outlier cluster randomly\\
                                        & on other clusters.\\
  Use first and second moment of each   & Use median and MAD of clusters for  \\
  cluster as initial values for $ f$ and $\sigma.$ & $ f$ and $\sigma.$\\
  Choose $\Pi$ and $\X_0$ according to  & Choose $\Pi$ and $\X_0$ according to \\
  cluster probabilities.                & cluster probabilities.\\
  \hline
  \multicolumn{2}{l}{}\\
  \multicolumn{2}{l}{\textbf{E-step:} Determine RN-derivative and calculate E-Step.}\\
  \multicolumn{2}{l}{}\\
  \hline
  Find RN-derivative. $\Lambda$  & Robustified version of $\Lambda$ \\
   & through suitable clipped version of $\lambda_k.$\\
  Estimate recursive filters   & No further robustification needed; i.e.,\\
  $J_k^{ji},$ $O_{k}^i$ and $T_k^{i}(g).$ & take over
  $J_k^{ji},$ $O_{k}^i$ unchanged and skip $T_k^{i}(g).$\\
  \hline
  \multicolumn{2}{l}{}\\
  \multicolumn{2}{l}{\textbf{M-step 1:} Obtain estimates for $f$ and $\sigma.$ }\\
  \multicolumn{2}{l}{}\\
  \hline
  MLE-estimates for $f$ and $\sigma$ through & Likelihoods re-stated; they are expressed \\
  recursive filters $O_k^{i}$ and $T_k^{i}(g).$ & as weighted sums of the observations $y_k.$ \\
  Recursive filters are substituted into likelihood. &  Robustified version of MLE through asymptotic\\
   & linear estimators.\\
  Estimates updated after each batch. & Estimates updated after each batch, \\
   & recursivity cannot be preserved completely.\\
  \hline
  \multicolumn{2}{l}{}\\
  \multicolumn{2}{l}{\textbf{M-step 2:} Obtain ML-estimators $\Pibold.$}\\
  \multicolumn{2}{l}{}\\
  \hline
  MLE-estimation, recursive filters $J_k^{ji}$ and  & Robustification through robust version of filters. \\
  $O_k^i$ are substituted into likelihood. & $J_k^{ji}$ and $O_k^i,$ no further observation $y_k$ \\
                                                    & has to be considered. \\
  \hline
  \multicolumn{2}{l}{}\\
  \multicolumn{2}{l}{\textbf{Rec:} Algorithm runs on next batch.}\\
  \multicolumn{2}{l}{}\\
  \hline
  Go to (RN) to compute the next batch. & Go to (RN) to compute the next batch.\\
  \hline
\end{tabular}
\caption{Classical algorithm setting and robustified version of each step.}
\label{TabAlgo}
\end{table}
%-------------------------------------------------------------------------------
\subsection{Robustification of Step (0)}
%%-------------------------------------------------------------------------------
So far, little has been said as to the initialization even in the non-robustified
setting. Basically, all we have to do is to make sure that the EM algorithm converges.
In prior applications of this algorithms (\cite{M:E:G:08} and \cite{E:M:D:09} amongst others), one approach
was to fill $\Pi$ with entries $1/N$, i.e., with uniform (and hence non-informative)
distribution over all states, independent from state $\X_0$. As to $f_i$ and $\sigma_i$,
an ad hoc approach would estimate the global mean and variance over all states and
then, again in a non-informative way jitter the state-individual moments, adding
independent noise to it. In our preliminary experiments, it turned out that this
na\"ive approach could drastically fail in the presence of outliers, so we instead
propose a more sophisticated approach which can also be applied in a classical (i.e.,
non-robust) setting: In a first step we ignore the time dynamics and interpret
our observations as realizations of a Gaussian Mixture Model, for which we use
{\sf R} package \texttt{mclust} (\cite{Fr:Ra:02,F:R:M:S:12}) to identify the mixture
components, and for each of these, we individually determine the moments $f_i$ and
$\sigma_i$. As to $\Pi$, we again assume independence of $\X_0$, but fill the
columns according to the estimated frequencies of the mixture components. In case
of the non-robust setting we would use $N$ mixture components and for each of them
determine $f_i$ and $\sigma_i$ by their ML estimators (assuming independent observations).
For a robust approach, we use $N+1$ mixture components, one of them---the one with the
lowest frequency---being a pure noise component capturing outliers. For each non-noise
component we retain the ML estimates for $f_i$  and $\sigma_i$. The noise
component is then randomly distributed amongst the remaining components, respecting
their relative frequencies prior to this redistribution. 

We are aware of the fact that reassigning the putative outliers at random could
be misleading in ideal situations (with no outliers) where one cluster could be split 
off into two but not necessarily so. Then in our strategy, the smaller offspring of 
this cluster would in part be reassigned to wrong other clusters, so this could still 
be worked on. On the other hand, this choice often works reasonably well, and as 
more sophisticated strategies are questions for model selection, we defer them
to further work.

Based on the $f_i$  and $\sigma_i$, for each observation $j$ and each state $i$, we get 
weights $0\leq w_{i,j}\leq 1$, $\sum_i w_{i,j}=1$ for each $j$, representing the likelihood 
that observation $j$ is in state $i$. For each $i$, again we determine robustified moment 
estimators $f_i'$, $\sigma_i'$ as weighted medians and scaled weighted MADs (medians of 
absolute deviations).
\medskip

\textbf{Weighted Medians And MADs:}
For weights $w_j\ge0$
and observations $y_j$, the weighted median $m=m(y,w)$ is defined as
$m={\rm argmin}_f \sum_j w_{j} |y_j-f|$, and with $y'_{j}=|y_j-m|$, the scaled weighted MAD $s=s(y,w)$
is defined as $s= c^{-1} {\rm argmin}_t  \sum_j w_{j} |y'_{j}-t|$, where $c$ is a consistency
factor to warrant consistent estimation of $\sigma$ in case of Gaussian observations,
i.e., $c ={\rm argmin}_t  \Ew \sum_j w_{j} \big||\tilde y_{j}|-t\big|$ for
$\tilde y_{j}\iids{\cal N}(0,1)$. $c$ can be obtained empirically for a sufficiently
large sample size $M$, e.g., $M=10000$, setting $c=\frac{1}{M}\sum_{k=1}^M c_{k}$,
$c_{k} ={\rm argmin}_t  \sum_j w_{j} \big||y''_{j,k}|-t\big|$, $y''_{j,k}\iids{\cal N}(0,1)$.\\

As to the (finite sample) breakdown point $FSBP$ of the weighted median (and
at the same time for the scaled weighted MAD), we define
$w_{j}^0=w_{i,j}/ \sum_{j'} w_{j'}$, and for each $i$ define the ordered weights
$w_{(j)}^0$ such that $w_{(1)}^0 \ge w_{(2)}^0 \ge \ldots \ge w_{(k)}^0$.
Then the FSBP in both cases is
$k^{-1} \min\{j_0=1,\ldots,k\,\mid\, \sum_{j=1}^{j_0} w_{(j_0)}^0 \ge k/2 \}$
which (for equivariant estimators) can be shown to be the largest possible value.
So using weighted medians and MADs, we achieve a decent degree of robustness
against outliers. E.g., assume we have $10$ observations with weights
$5 \times 0.05; 3 \times 0.1; 0.2; 0.25$. Then we need at least three outliers
(placed at weights $0.1,0.2,0.25$, respectively) to produce a breakdown.

%-------------------------------------------------------------------------------
\subsection{Robustification of the E-step}
%%-------------------------------------------------------------------------------
As indicated, in this step we cannot recur to asymptotics, but rather have
to appeal to a theory particularly suited for this recursive setting.
In particular, the SO-neighborhoods introduced in \eqref{YSO} turn out
to be helpful here.
%%-------------------------------------------------------------------------------
\subsubsection{Crucial Optimality Thm}
%%-------------------------------------------------------------------------------
%
Consider the following optimization problem of reconstructing the ideal observation $Y^{\ssr id}$
by means of the realistic/possibly contaminated $Y^{\ssr re}$ on an SO-neighborhood.
\paragraph{Minimax-SO problem}
Minimize the maximal MSE on an SO-neighborhood, i.e., find a $Y^{\ssr re}$-measurable reconstruction $f_0$
for $Y^{\ssr id}$ s.t.\
\begin{align}
\quad&\max\nolimits_{{\cal U}}\, \Ew_{\SSs\rm re} |Y^{\ssr id}-f(Y^{\rm\SSs re})|^2 = \min\nolimits_f{}! \label{minmaxSO}
\end{align}

The solution is given by
\begin{Thm}[Minimax-SO]\label{ThmSO} %\mbox{\hspace{1mm}}\\[-1.5ex]
In this situation, there is a \emph{saddle-point\/} $(f_0, P_0^{Y^{\rm\SSs di}})$ for Problem~\eqref{minmaxSO}
\begin{eqnarray}
%\quad
f_0(y)\!\!&\!\!:=\!\!&\!\!\Ew Y^{\ssr id} +H_\rho(D(y)),\qquad H_b(z)=z \min\{1, b/|z|\} \label{f0def}\\
%\quad
P_0^{Y^{\rm\SSs di}}(dy)\!\!&:=\!\!&\!\!\Tfrac{1-r}{r} ( \big|D(y)\big|\!/\!\rho\,-1)_{\SSs +}\,\, P^{Y^{\rm\SSs id}}(dy) \label{P0def}
\end{eqnarray}
where $\rho>0$ ensures that $\int \,P_0^{Y^{\rm\SSs di}}(dy)=1$ and
\begin{equation} \label{Ddef}
D(y)=y-\Ew Y^{\ssr id}
\end{equation}
The value of the minimax risk of Problem~\eqref{minmaxSO} is
\begin{equation} \label{sadvalSO}
\tr \Cov (Y^{\ssr id}) -(1-r)\Ew_{\rm\SSs id}\big[\min\{|D(Y^{\rm\SSs id})|, \rho\}^2\, \big]
\end{equation}
\end{Thm}
\begin{proof}{}
See Appendix \ref{app}.
\end{proof}
The optimal procedure in equation~\eqref{f0def} has an appealing interpretation:
It is a compromise between the (unobservable) situations that (a) one observes
the ideal $Y^{\rm\SSs id}$, in which case one would use it unchanged and (b) one
observes $Y^{\rm\SSs di}$, i.e.; something completely unrelated to $Y^{\rm\SSs id}$,
hence one would use the best prediction for $Y^{\rm\SSs id}$ (in MSE-sense) without
any information, hence the unconditional expectation $\Ew Y^{\ssr id}$. The decision
on how much to tend to case (a) and how much to case (b) is taken according to the
(length of the) discrepancy $D(Y^{\rm\SSs re})$ between observed signal
$Y^{\rm\SSs re}$ and $\Ew Y^{\ssr id}$. If this length is smaller than $\rho$,
we keep $Y^{\rm\SSs re}$ unchanged, otherwise we modify $\Ew Y^{\ssr id}$ by adding
a clipped version of $D(Y^{\rm\SSs re})$.

%%-------------------------------------------------------------------------------
\subsubsection{Robustification of Steps (RN), (E)}
%%-------------------------------------------------------------------------------
%
In the Girsanov-/change of measure step we recall that the
corresponding likelihood ratio here is just
\begin{equation}
\lambda_s:=\frac{\sigma^{-1}(\X_{s-1})\varphi\Big(\sigma^{-1}(\X_{s-1})\big(y_s- f(\X_{s-1})\big) \Big)}{\varphi(y_s)}
\end{equation}
Apparently $\lambda_s$ can both take values close to $0$, and, more dangerously, in particular for small values
of $\sigma(\X_{s-1})$, very large values. So bounding the $\lambda_s$ is crucial to avoid effects like
in Figure~\ref{MSCIRetOut2}.

A first (non-robust) remedy uses a data driven reference measure, i.e., instead of ${\cal N}(0,1)$, we use
${\cal N}(0,\bar \sigma^2)$ where $\bar \sigma$ is a global scale measure taken over all observations,
ignoring time dependence and state-varying $\sigma$'s. A robust proposal would take $\bar \sigma$ to be
the MAD of all observations (tuned for consistency at the normal distribution). This leads to
\begin{equation}
\tilde \lambda_s:=\frac{\sigma^{-1}(\X_{s-1})\varphi\Big(\sigma^{-1}(\X_{s-1})\big(y_s- f(\X_{s-1})\big) \Big)}{\bar\sigma^{-1}\varphi(\bar\sigma^{-1} y_s)}
\end{equation}
Eventually, in both estimation and filtering/prediction, $\bar\sigma$ cancels out as a common factor
in nominator and denominator, so is irrelevant in the subsequent steps; its mere purpose is to
stabilize the terms in numeric aspects.

To take into account time dynamics in our robustification, we want to use Theorem~\ref{ThmSO}, but to this
end, we need second moments, which for $\lambda_s$ need not exist. So instead, we apply the theorem
to $Y^{\rm\SSs id}=\sqrt{\tilde\lambda_s}$, which means that $\tilde\lambda_s=(Y^{\rm\SSs id})^2$  is robustified by
\begin{equation}
\bar \lambda_s = (\Ew_{\rm\SSs id} \sqrt{\tilde\lambda_s} +
H_b(\sqrt{\tilde\lambda_s}-\Ew_{\rm\SSs id} \sqrt{\tilde\lambda_s}))^2
\end{equation} for $H_b(x)=x\min\{1,b/|x|\}$. Clipping height $b$ in turn is chosen such that
$\Ew \bar \lambda_s=\alpha$, $\alpha=0.95$ for instance. As in the ideal situation $\Ew \lambda_s=1$,
in a last step with a consistency factor $c'_s$ determined similarly to $c_i$ in the initialization
step for the weighted MADs, we pass over to $\bar \lambda^0_s=c_s \bar \lambda_s$
such that $\Ew \bar \lambda^0_s=1$.\medskip\\

Similarly, in the remaining parts of the E-step, for each of the filtered processes generically
denoted by $G$ and the filtered one by $\hat G$, we could replace $\hat G$ by
\begin{equation}
\bar G = \Ew_{\rm\SSs id} \hat G+ H_b(\hat G-\Ew_{\rm\SSs id} \hat G)
\end{equation}
for $G$ any of $J_k^{ji}$, $O_k^{i}$, and $T_k^{i}(f)$ and again suitably chosen $b$.

It turns out though, that it is preferable to pursue another route.
The aggregates $T_k^{i}(f)$ are used in the M-step in \eqref{llik},
but for a robustification of this step, it is crucial to be able to
attribute individual influence to each of the observations, so instead we split
up the terms of the filtered neg-loglikelihood into summands
$w_{i,j}/ \hat O_k^i [(y_j-f_i)^2/\sigma_i^2+\log \sigma_i]$ for $j=1,\ldots,k$,
and hence we may skip a robustification of $T_k^{i}(f)$. Similarly, as $J_t^{ji}$, $O_t^{i}$
are filtered observations of multinomial-like variables, a robustification is of limited use,
as any contribution of a single observation to these variables can at most be
of absolute value $1$, so is bounded anyway. Hence in the E-step, we only
robustify $\lambda_s$.
The splitting up the aggregates $T_k^{i}(f)$ into summands amounts to
giving up strict recursivity, as for $k$ observations in one batch,
one now has to store the values $w_{i,j}/ \hat O_k^i$ for $j=1,\ldots,k$,
and building up from $j=1$, at observation time $j=j_0$ within the batch,
we construct $w_{i,j;j_0}$, $j=1,\ldots,j_0$ from the values $w_{i,j;j_0-1}$,
$j=1,\ldots,j_0-1$, so we have a growing triangle of weight values.
This would lead to increasing memory requirements, if we had not chosen
to work in batches of fixed length $k$, which puts an upper bound onto
memory needs.

\subsection{Robustification of the (M)-Step}
%%-------------------------------------------------------------------------------
As mentioned before, contrast to the (E)-step, in this estimation step,
we may work with classical gross error neighborhoods~\eqref{GEN} and with the standard
i.i.d. setting.
%%-------------------------------------------------------------------------------
\subsubsection{Shrinking Neighborhood Approach}
%%-------------------------------------------------------------------------------
By Bienaymé, variance then usually is $\LO(1/n)$ for sample size $n$,
while for robust estimators, the maximal bias is proportional to
the neighborhood radius $\ve$. Hence unless $\ve$ is appropriately scaled in $n$,
for growing $n$, bias will dominate eventually for growing $n$. This is avoided in the shrinking
neighborhood approach by setting $\ve=\ve_n=r/\sqrt{n}$ for some $r \ge 0$,
compare \cite{Ried:94}. \cite{K:R:R:09} sets $\ve=\ve_n=r/\sqrt{n}$
for some initial radius $r\in [0,\infty)$. One could see this shrinking
as indicating that with growing $n$, diligence is increasing so the
rate of outliers is decreasing. This is perhaps overly optimistic. Another
interpretation is that the severeness of the robustness problem with $10\%$
outliers at sample size $100$ should not be compared with the one
with $10\%$ outliers at sample size $10000$ but rather with the one with
$1\%$ outliers at this sample size.

In this shrinking neighborhood setting, with
mathematical rigor, optimization of the robust criteria can
be deferred to the respective IFs, i.e. instead of determining
the IF of a given procedure, we construct a procedure to a given
(optimally-robust) IF.

This is achieved by the concept of \textit{asymptotically linear estimators} (ALEs),
as it arises canonically in most proofs of asymptotic normality:
In a smooth ($L_2$-differentiable) parametric model  ${\cal P}=\{P_\theta,\;\;\theta\in\Theta\}$ for
i.i.d observations $X_i\sim P_\theta$ with
open parameter domain $\Theta\subset\R^d$ based on the scores\footnote{Usually
$\Lambda_\theta$ is the logarithmic derivative of the density w.r.t.\
the parameter, i.e., $\Lambda_\theta(x)=\partial/\partial \theta
\log p_\theta(x)$.} $\Lambda_\theta$ and its finite Fisher information
${\cal I}_\theta=\Ew_\theta \Lambda_\theta\Lambda_\theta^\tau$,
we define the set $\Psi_2(\theta)$ of influence functions as the
subset of $L^d_2(P_\theta)$ consisting of square integrable functions
$\psi_\theta$ with $d$ coordinates with $\Ew_\theta \psi_\theta=0$ and
$\Ew_\theta \psi_\theta \Lambda_\theta^\tau= \EM_d$ where $\EM_d$ is the
$d$-dimensional unit matrix. Then a sequence of estimators $S_n=S_n(X_1,\ldots,X_n)$
is called an ALE if
\begin{equation} \label{ALE}
S_n=\theta+\frac{1}{n}\sum_{i=1}^n \psi_\theta(X_i) +
   \Lo_{P_\theta^n}(n^{-1/2})
\end{equation}
for some influence function $\psi_\theta\in \Psi_2(\theta)$.
In the sequel we fix the true $\theta\in\Theta$
and suppress it from notation where clear from context.

In particular, the MLE usually has influence function
$\psi^{\rm\SSs MLE}={\cal I}^{-1} \Lambda$, while most other common
estimators also have a representation~\eqref{ALE} with a different
$\psi$.

For given IF $\psi$ we may construct an ALE $\hat \theta_n$ with $\psi$ as IF by
a one-step construction, often called one-step-reweighting: To given
starting estimator $\theta^0_n$ such that
$R_n^0=\theta^0_n-\theta=\Lo_{P^n_\theta}(n^{-1/4+{0}})$ we define
\begin{equation} \label{onest}
\hat \theta_n=\theta^0_n+ \frac{1}{n}\sum_{j=1}^n \psi_{\theta^0_n}(X_j)
\end{equation}
Then indeed
$\hat \theta_n=\theta+\frac{1}{n}\sum_{j=1}^n \psi_{\theta}(X_j)+R_n
$ and $R_n=\Lo_{P^n_\theta}(n^{-1/2})$, i.e., $\hat \theta_n$ forgets about $\theta^0_n$
as to its asymptotic variance and GES; however its breakdown point
is inherited from $\theta^0_n$ once $\Theta$ is unbounded and $\psi$ is bounded.
Hence for the starting estimator, we seek for  $\theta^0_n$ with high breakdown
point.

For a more detailed account on this approach, see \cite{Ried:94}.
%%-------------------------------------------------------------------------------
\subsubsection{Shrinking Neighborhood Approach With Weighted Observations}
%%-------------------------------------------------------------------------------
As mentioned, coming from the E-step, not all observations $y_j$ are equally
likely to contribute to state $i$, hence we are in a situation with weighted
observations, where we may pass over to normed weights $w_{i,j}^0=w_{i,j}/ \sum_{j'} w_{i,j'}$
summing up to $1$.

\begin{figure}
\begin{center}
\includegraphics[scale=0.6]{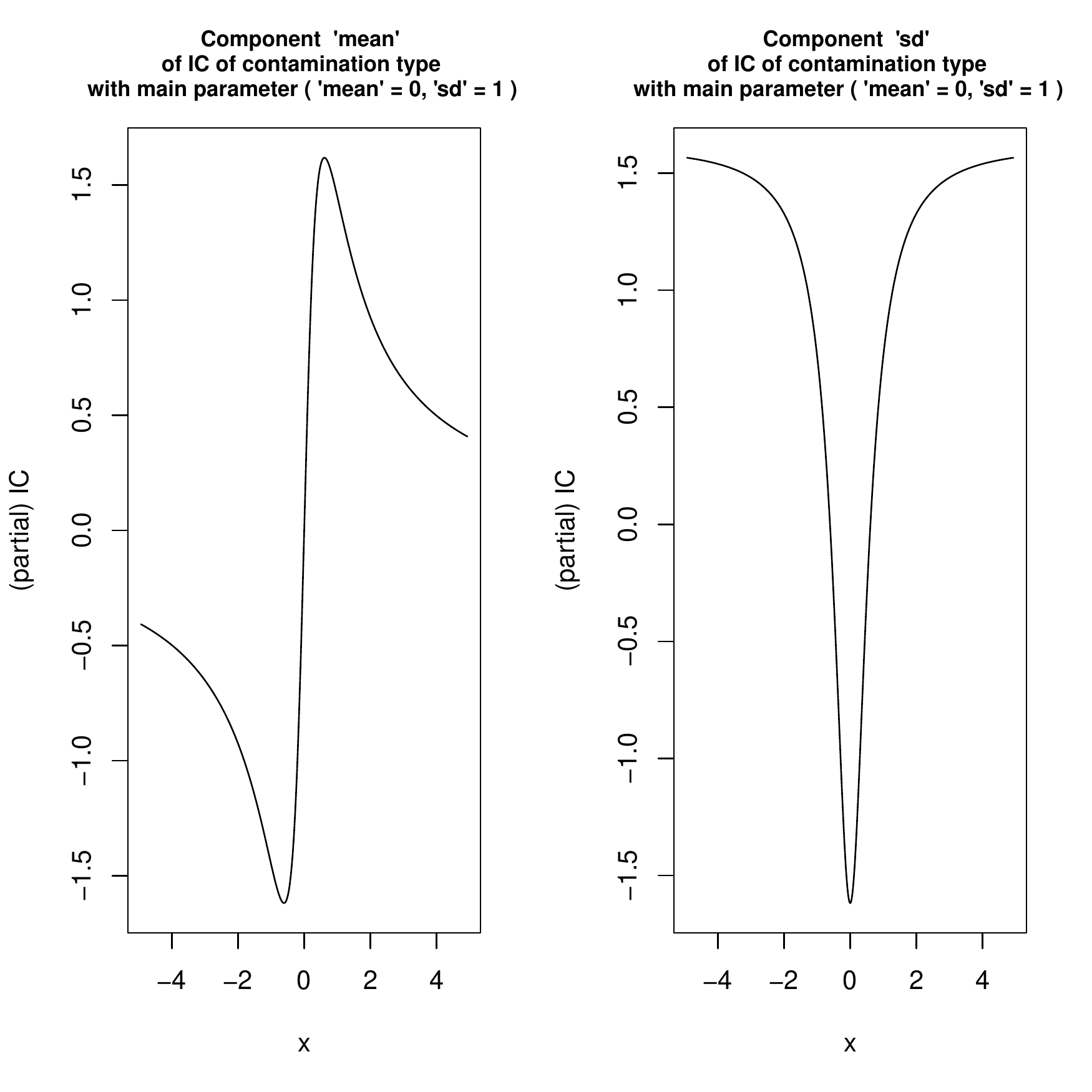}
\caption{Influence function of the MBRE at ${\cal N}(0,1)$; left panel: location part; right panel: scale part.}
\label{MBRE.pdf}
\end{center}
\end{figure}

Suppressing state index $i$ from notation here, with these weights and $\theta^0_n$ the vector of weighted median and scaled
weighted MAD for state $i$, \eqref{onest} becomes
\begin{equation} \label{onestweight}
\hat \theta_n=\theta^0_n+ \sum_{j=1}^n w_{j}^0 \psi_{\theta^0_n}(y_j)
\end{equation}
for $\hat \theta_n$ again a two-dimensional ALE with location and scale coordinate
and $\psi_{\theta}(y)= \sigma \psi((y-f)/\sigma)$, at $\theta=(f,\sigma)^\tau$, is the
IF  of the MBRE in the one-dimensional Gaussian location and scale model at ${\cal N}(0,1)$;
i.e.,
\begin{equation} \label{MBRE1}
\psi(y)= b Y(y) / |Y(y)|,\qquad Y(y)=(y,A(y^2-1)-a),
\end{equation}
with numerical values for $A,a,b$ up
to four digits  taken from {\sf R} package
\texttt{RobLox}, \cite{RobLox}, being
\begin{equation} \label{MBRE2}
A=0.7917, \qquad a=-0.4970,\qquad b=1.8546,
\end{equation}
Influence function $\psi$ is illustrated in Figure~\ref{MBRE.pdf}.

To warrant positivity of $\sigma$ and to maintain a high breakdown point even in the
presence of inliers (driving $\sigma$ to essentially $0$), for the scale component,
instead of \eqref{onestweight} we use the asymptotically equivalent form
\begin{equation}
\hat \sigma_n=\sigma^0_n \exp\Big[\sum_{j=1}^n w_{j}^0 \psi_{\rm scale}\big(\,(y_j-\mu^0_n)/\sigma^0_n\,\big)\Big].
\end{equation}
Using the MBRE-$\psi$ from \eqref{MBRE1} and \eqref{MBRE2} on first glance
could be seen as overly cautious. Detailed simulation studies, compare
e.g.\ \citet{K:D:10}, show that for our typical batch lengths of 10--20,
the MBRE also is near to optimal in the sense of \citet{R:K:R:08} in the
situation where nothing is known about the true outlier rate (including, of course the situation
where no outliers at all occur).
\subsubsection{Robustification of Steps (M1) \& (M2)}
%%-------------------------------------------------------------------------------
Now, we derive robust estimators of the model parameters $f_i$ and $\sigma_i$,
i.e., we justify passage to weighted ALEs as in \eqref{onestweight}. In particular
we specify the weights $w_{j}^0=w_{i,j}^0$ therein.

 Recall, that the M1-step in the classical algorithm gives the optimal parameter estimates stated in Theorem \ref{theoremParamEst}. We now build ALEs, which can be achieved, when the MLEs of the parameters $f_i$ and $\sigma_i$ are stated as weighted sums of the observations $y_k.$
\begin{Thm}
\label{optParamWeightedSum}
With
\begin{equation}
w_{i,l}^0= \frac{\langle \hat{\X}_{l-1}, \e_i\rangle}{\eta (O^{(i)})_k}
\end{equation}
the optimal parameter estimates $\hat{f}_{i}$ and $
\hat{\sigma}_i $ are given by
\begin{eqnarray}
\label{fhatRob}
\widehat f_i &=& \sum_{l=1}^k w_{i,l}^0 y_l\\
\label{sigmahatRob}
 \widehat \sigma_i^2&=&\sum_{l=1}^k w_{i,l}^0 (y_l-f_i)^2.
\end{eqnarray}
\end{Thm}
\begin{proof}{}
To find the optimal estimate for $f$ consider
\begin{eqnarray*}
\Lambda_k^* &:=&\prod_{l=1}^k \lambda_l^*\,\,\mbox{ with }
\lambda_l^*:=\exp\Bigl(\frac{(y_l-\langle f,\X_{l-1}\rangle)^2-(y_l-\langle \widehat{f},\X_{l-1}\rangle)^2}{2\langle \sigma,\X_{l-1}\rangle^2}\Bigr).
\end{eqnarray*}
Up to constants irrelevant for optimization, the filtered log-likelihood is then
\begin{equation}\label{llikw1}
\E[\ln(\Lambda_k^*)\mid {\cal F}_k^y]=\sum_{i=1}^N\sum_{l=1}^k w_{i,l}^0 (y_l-\widehat{f}_i)^2
\end{equation}
Maximising the log-likelihood $\E[ln(\Lambda_k^*)\mid {\cal F}_k^Y]$ in $\widehat{f}_i$ hence leads to the optimal parameter estimate
\begin{eqnarray*}
\sum_{l=1}^k  \langle {\X}_{l-1}, \e_i\rangle (2y_l\widehat{f}_i-\widehat{f}_i^2)=0 \qquad \Longrightarrow\qquad \widehat{f}_i(k)
=\sum_l^k w_{i,l}^0 y_l
\end{eqnarray*}
In an analogue way, for $\sigma_i$ we define
\begin{eqnarray*}
\Lambda_k^{+}&:=&\prod_{l=1}^k \lambda_l^{+}\,\,\mbox{ with }
\lambda_l^{+}:=\frac{\langle \sigma, \X_{l-1}\rangle}{\langle\widehat{\sigma}, \X_{l-1}\rangle}\exp\Bigl(\frac{(y_l-\langle f,\X_{l-1}\rangle)^2}{2\langle \sigma,\X_{l-1}\rangle^2}-\frac{(y_l-\langle f,\X_{l-1}\rangle)^2}{2\langle \widehat{\sigma},\X_{l-1}\rangle^2}\Bigr).
\end{eqnarray*}
and hence, again up to irrelevant terms
\begin{equation} \label{llikw}
\E[\ln(\Lambda_k^+)\mid {\cal F}_k^y]=\sum_{i=1}^N\sum_{l=1}^k \langle \hat{\X}_{l-1}, \e_i\rangle [\frac{(y_l-\widehat{f}_i)^2}{2\sigma_i^2}+ \log(\sigma_i)]
\end{equation}
From this term, which has to be minimised for $\widehat{\sigma}_i$ we get
\begin{eqnarray*}
\sum_{l=1}^k \langle {\X}_{l-1}, \e_i\rangle \Bigl[ -\frac{1}{2\widehat{\sigma}^3}(y_l-\widehat{f}_i)^2+\frac{1}{\widehat{\sigma}_i}\Bigr]=0
\qquad\Longrightarrow\qquad \widehat{\sigma}_i^2=\sum_{l=1}^k w_{i,l}^0 (y_l-\widehat{f}_i)^2
\end{eqnarray*}
Note that \eqref{llikw} takes its minimum at the same place as
\begin{equation} \label{llikw2}
\ln(\Lambda_k^{++})=\sum_{i=1}^N\sum_{l=1}^k w_{i,l}^0 [(y_l-\widehat{f}_i)^2-\sigma_i^2]^2
\end{equation}
\end{proof}
For the robustification of the parameter estimation (step M1) we now distinguish two approaches. The first robustification is utilized in the first run over the first batch of data and is therefore called the \textit{initialization} step M1. The robust estimates of the parameters from the second batch onwards are then achieved through a weighted ALE.
\begin{Thm}
The robust parameter estimates for the model parameters $f_i$ and $\sigma_i$ in the (1) initialization and (2) all following batches are given by
\begin{enumerate}
\item Replacing, for initialization, the squares by absolute values in \eqref{llikw1} and in \eqref{llikw2},
$\hat{f}_i$ and $\hat\sigma_i$  are the weighted median and scaled weighted MAD, respectively, of
the $y_l$, $l=1,\ldots,k$ with weights $(w_{i,l}^0)_l$.
\item For further batches, the weighted MBRE is obtained as a one-step construction with the parameter estimate
 $(f_i^0,\sigma_i^0)$ from the previous batch as starting estimator and with IF $\psi=(\psi_{\rm loc},\psi_{\rm scale})$ from
 \eqref{MBRE1}, \eqref{MBRE2}, i.e.,
\begin{eqnarray}
\hat{f}_i&=&f_i^0+\sigma_i^0\sum_{l=1}^k w_{i,l}^0\psi_{\rm loc}\bigl((y_l-f_i^0)/\sigma_i^0\bigr)\\
\hat{\sigma}_i&=&\sigma_i^0 \exp\Big(\sum_{l=1}^k w_{i,l}^0\psi_{\rm scale}\bigl((y_l-f_i^0)/\sigma_i^0\bigr)\Big)
\end{eqnarray}
\end{enumerate}
\end{Thm}
\begin{proof}
\ \\
\begin{enumerate}
\item Initialization:
With absolute values instead of squares, \eqref{llikw1} becomes
$$\hat{f}_i=\argmin_{f_i}\sum_{l=1}^k w_{i,l}^0|y_l-f_i|$$
Now if $w_{i,l}^0$ is constant in $l$, this leads to the empirical median as unique minimizer justifying the name.
For the scaled weighted MAD, the argument parallels the previous one, leading to consistency factor $c_i=\Phi(3/4)$ for $\Phi$ the cdf of ${\cal N}(0,1)$.
\item M1 in further batches:
Apparently, by definition, $(\hat f_i, \hat \sigma_i)$ is an ALE, once we show that $\psi$ is square integrable, $\Ew(\psi)=0$, $\Ew(\psi \Lambda')= \EM_2$. The latter two properties can be checked numerically, while by boundedness square integrability is obvious. In addition it has the
necessary form of an MBRE in the i.i.d. setting as given in \cite[Thm~5.5.1]{Ried:94}. To show that this also gives the MBRE in the context of weighted observations, we would need to develop the theory of ALEs for triangular schemes similar to the one in the Lindeberg Feller Theorem. This has been done, to some extent in \cite[Section~9]{Ruc:01}. In particular, for each state $i$, we have to assume a Noether condition excluding
observations overly influential for parameter estimation in this particular state, i.e.,
\begin{equation}
\lim_{k\to\infty} \max_{l=1,\ldots,k} (w_{i,l;k}^0)^2/\sum_{j=1}^k (w_{i,l;k}^0)^2 = 0
\end{equation}
We do not work this out in detail here, though.
\end{enumerate}
\end{proof}

Consider again our filtering algorithm and recall, that the filter runs over the data set in batches of roughly ten two fifty data point. To determine the ALE for our parameters, we have to calculate the weights $w_{i,l}^0= \langle \widehat{\X}_{l-1}, \e_i\rangle/\widehat{O}_k^i.$ Therefore, our algorithm has to know all values of $\hat\X_l$ from $1$ to $k$ in each batch. With this, our robustification of the algorithm cannot obtain the same recursiveness as the classical algorithm. However, since we only have to determine and save the estimates of $\X_l$ in each batch, the algorithm still is numerically efficient, the additional costs are low.
In general, the ALEs are fastly computed robust estimators, which lead in our case to a fast and, over batches, recursive algorithm.

The additional computational burden to store all the weights $w_{i,l}^0$ arising in the
robustification of the M1-step is more than paid off by the additional benefits they offer for
diagnostic purposes beyond the mere EM-algorithm:
They tell us which of the observations, due to their likelihood to be in state $i$ carry more information on the respective
parameters $f_i$ and $\sigma_i$ than others. The same goes for the terms $\psi_{\theta}(y_l)$ which capture the individual
information of observation $y_l$ for the respective parameters. Even more though, the coordinates of
$\psi_{\theta}(y_j)/|\psi_{\theta}(y_l)|$ tell us how much of the information in observation $y_l$ is used for estimating
$f_i$ and how much for $\sigma_i$. In addition the function $y\mapsto w_{i,l}^0 \psi_{\theta}(y)$ can be used for sensitivity analysis,
telling us what happens to the parameter estimates for small changes in observation $y$. Finally, using the unclipped, classically
optimal IF of the MLE, but evaluated at the robustly estimated parameters, we may identify outliers not fitting to the ``usual'' states.
\section{Implementation and Simulation}
\label{Sec_Implementation}
The classical algorithm as well as the robust version are implemented in {\sf R};
we plan to release the code in form of a contributed package on
\href{http://cran.r-project.org/}{\tt CRAN} at a later stage. The implementation
builds up on, respectively uses contributed packages {\tt RobLox} and {\tt mclust}.
At the time of writing we are preparing a thorough simulation study to explore
our procedure in detail and in a quantitative way. For the moment, we restrict
ourselves to assess the procedure in a qualitative way, illustrating how it
can cope with a situation like in Figure~\ref{MSCIRetOut2}. 

\begin{figure}
\begin{center}
\includegraphics[height=6cm]{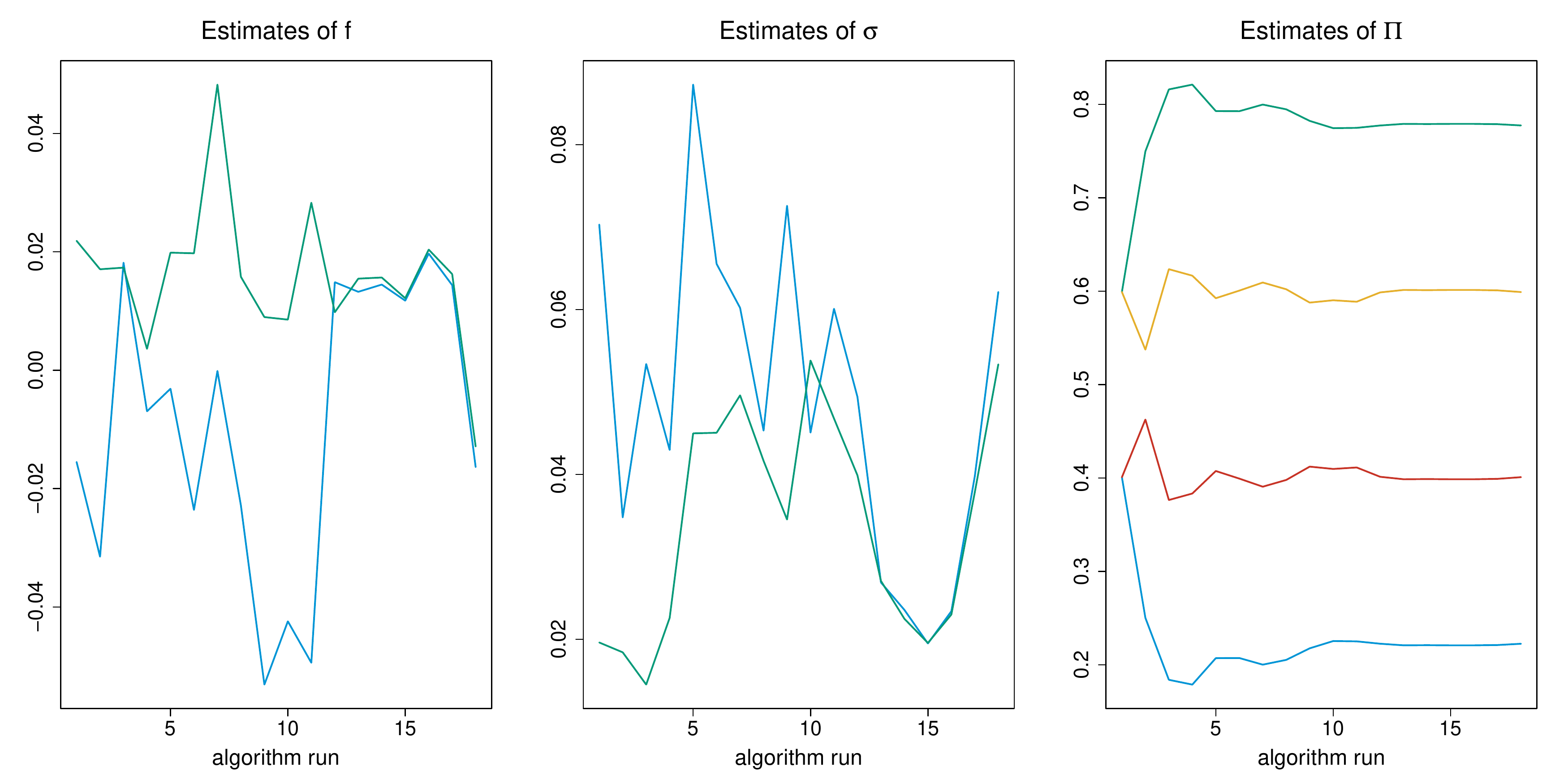}
\caption{Robust parameter estimates for monthly MSCI returns between 1994 and 2009---analogue to Figure~\ref{MSCIRet}}
\label{MSCIRet.rob}
\end{center}
\end{figure}

In Figure~\ref{MSCIRet.rob},
we see the paths of the robust parameter estimates for $f$, $\sigma$, and $\Pi$;
due to the new initialization procedure, the estimates---in particular those for $\Pi$---differ
a little from those of Figure~\ref{MSCIRet}. Still, all the estimators behave very reasonable
and are not too far from the classical ones. 

\begin{figure}
\begin{center}
\includegraphics[height=5cm]{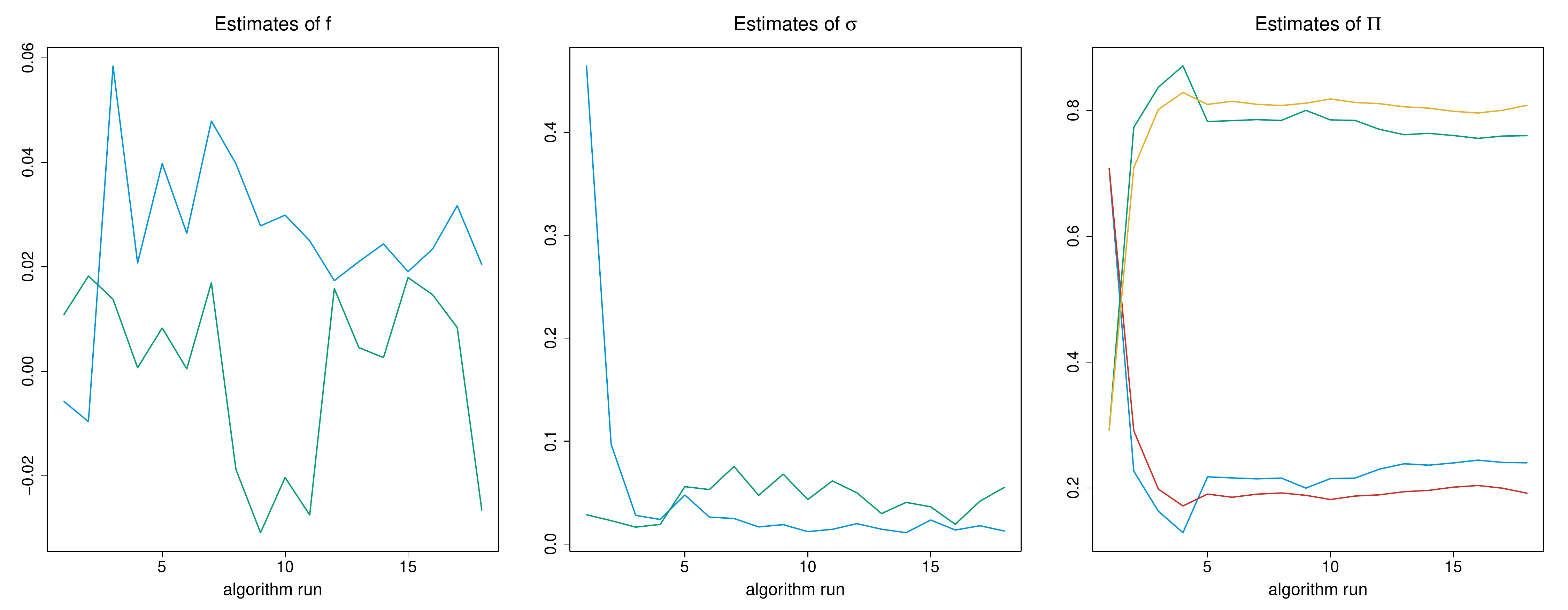}
\includegraphics[height=5cm]{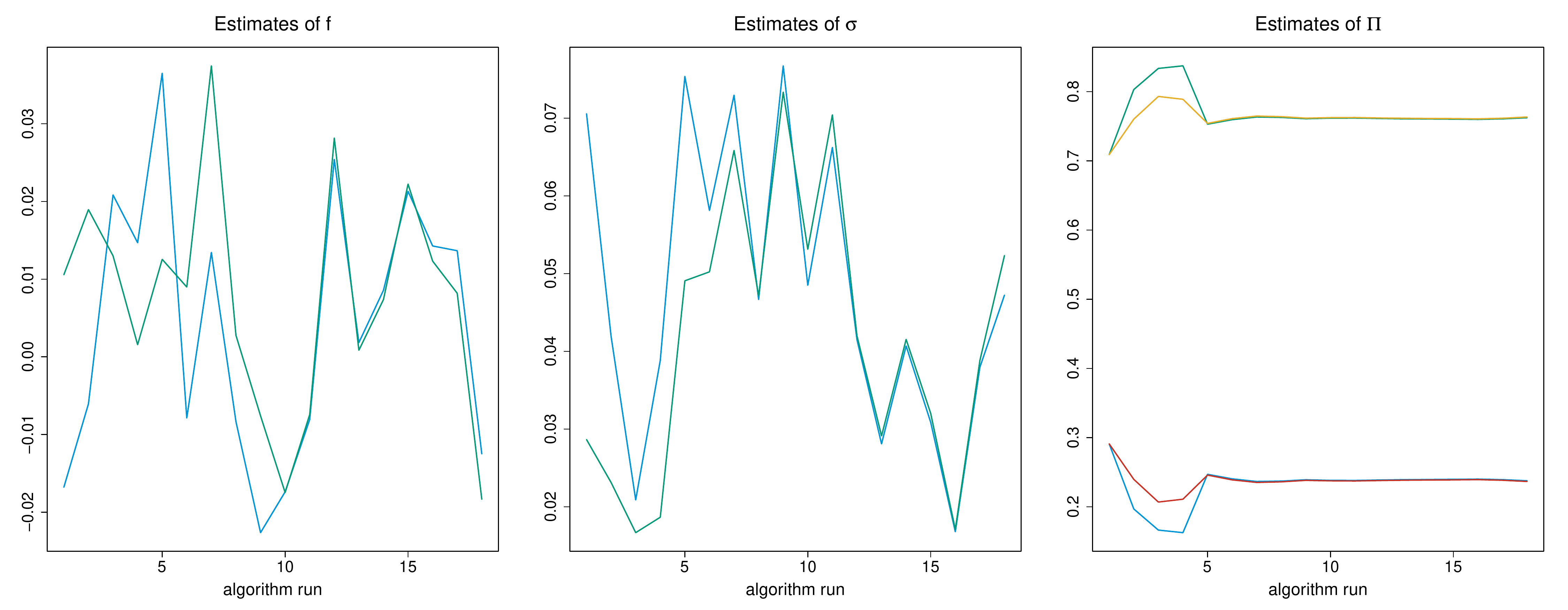}
\caption{Robust parameter estimates for monthly MSCI with planted outliers---analogue to Figures~\ref{MSCIRetOut1} and \ref{MSCIRetOut2}}
\label{MSCIRet2.rob}
\end{center}
\end{figure}

In the outlier situations from Figures~\ref{MSCIRetOut1}
and \ref{MSCIRetOut2}, illustrated in Figures~\ref{MSCIRet2.rob}, the estimates for
$f$ and $\sigma$ remain stable at large as desired. The estimates for $\Pi$ however do
get irritated, essentially flagging out one state as outlier state. Some more work remains
to be done to better understand this and to see how to avoid this.

Aside from this, our algorithm already achieves its goals; in particular, our procedure never breaks 
down---contrary to the classical one.

%% ------------------------------------------------------------------------
%
\section{Conclusion}
\label{Sec_Conclusion}
%% ------------------------------------------------------------------------
In financial applications, we often have to consider the case of outliers in our data set,
which can occur from time to time e.g., due to either wrong values in the financial database
or unusual peaks or lows in volatile markets. Conventional parameter estimation methods cannot
handle these specific data characteristics well.

\paragraph{Contribution of this paper:}
Our contribution to this issue is two fold:

First, we analyse step by step the general filter-based
EM-algorithm for HMMs by \cite{E:94} and highlight, which problems can occur in case of extreme values.
We extend the classical algorithm by a new technique to find initial values, taking into account the
$N-$state setting of the HMM. In addition, for numerical reasons we use a data-driven
reference measure instead of the standard normal distribution.

Second, we have proposed a full robustification of the classical EM-algorithm.
Our robustified algorithm is stable w.r.t.\ outliers in the observation process
and is still able to estimate processes of the Markov chain as well as optimal
parameter estimates with acceptable accuracy. The robustification builds up on concepts
from robust statistics like SO-optimal filtering and asymptotic linear estimators.
Due to the non-iid nature of the observations as apparent from the non-uniform
weights $w_{i,l}^0$ attributed to the observations, these concepts had to be
generalized for this situation, leading to weighted medians, weighted MADs, weighted ALEs.
Similarly, the SO-optimal filtering (with focus on state reconstruction) is not directly
applicable for robustifying the Radon-Nikodym terms $\lambda_s$, where we (a) had to
clean the ``observations'' themselves and (b) had to pass over to $\sqrt{\lambda_s}$
for integrability reasons.

Our robust algorithm is computationally efficient. Although complete recursivity cannot
be obtained, the algorithm runs over batches and keeps its recursivity there additionally
storing the filtered values of the Markov chain. This additional burden is outweighed though
by the benefits of these weights and influence function terms for diagnostic purposes.
As in the original algorithm, the model parameters, which are guided by the state of the
Markov chain, are updated after each batch, using a robust ALE however.
The robustification therefore keeps the characteristic of the algorithm, that new information,
which arises in the observation process, is included in the recent parameter update---there
is no forward-backward loop.

The forecasts of asset prices, which are obtained through the robustified parameter estimates,
can be utilized to make investment decisions in asset allocation problems.

To sum up, our forecasts are robust against additive outliers in the observation process and able to handle
switching regimes occurring in financial markets.\\

\paragraph{Outlook:}
It is pretty obvious how to generalize our robustification to a multivariate setting: The E-step is not affected by multivariate observations,
and the initialization technique using Gaussian Mixture Models ideas already is available in multivariate settings.
Respective robust multivariate scale and location estimators for weighted situations still have to be implemented, though,
a candidate being a weighted variant of the (fast) MCD-estimator, compare \citet{R:L:87,R:vD:99}.

 Future work will hence translate our robustification to a multivariate setting to directly apply the algorithm to asset allocation problems for portfolio optimisation. Furthermore, investment strategies shall be examined within this robust HMM setting to enable investors a view on their portfolio, which includes possible outliers or extreme events. The implementation of the algorithms shall be part of an {\sf R} package, including a thorough simulation study of the robustified algorithm and its application in portfolio optimisation.
 
Finally, an automatic selection criterion for the number of states to retain 
would be desirable which is a question of model selection, where criteria like 
BIC have still to be adopted for robustness.

%% ------------------------------------------------------------------------
\section*{Acknowledgement}
%% ------------------------------------------------------------------------
Financial support for C.~Erlwein from Deutsche Forschungsgemeinschaft (DFG)
within the project ``Regimeswitching in zeitstetigen Finanzmarktmodellen: Statistik und problemspezifische Modellwahl''
(RU-893/4-1) is gratefully acknowledged.
%% ------------------------------------------------------------------------
\appendix
\label{app}
%% ------------------------------------------------------------------------
\begin{small}
\section{Proofs}
\label{Sec_proofs}
\subsection*{Proof to Theorem~\ref{ThmSO}}
(1) Let us solve $\max_{\partial{\cal U}}\min_f{} [\ldots]$ first, which amounts to $\min_{\partial{\cal U}} \Ew_{\SSs \rm re}[\big|\Ew_{\SSs \rm re}[Y^{\rm\SSs id}|Y^{\rm\SSs re}]\big|^2]$.
For fixed element $P^{Y^{\rm\SSs di}}$ assume a dominating $\sigma$-finite measure $\mu$, i.e.,
 $\mu\gg P^{Y^{\rm\SSs di}}$, $\mu\gg P^{Y^{\rm\SSs id}}$; this gives us a $\mu$-density
$q(y)$ of $P^{Y^{\rm\SSs di}}$.
Determining the joint (real) law $P^{Y^{\rm\SSs id},Y^{\rm \SSs re}}(d\tilde y,dy)$ as
\begin{equation} \label{SOdistr}
P(Y^{\rm\SSs id} \!\in\! A, Y^{\rm \SSs re} \!\in\! B) =\!\! \int \!\!
\Jc_A(\tilde y)\Jc_B(y) [(1\!-\!r) \Jc(\tilde y=y) +r q(y)]\, p^{Y^{\rm\SSs id}}(\tilde y)\,\mu(d\tilde y)\mu(dy)
\end{equation}
we deduce that $\mu(dy)$-a.e.
\begin{equation} \label{SOEw}
\Ew_{\rm\SSs re}[Y^{\rm\SSs id}|Y^{\rm\SSs re}\!\!=\!y]=
\frac{r q(y)\!\Ew Y^{\rm\SSs id} \!+\!(1\!-\!r) y p^{Y^{\rm\SSs id}}(y)}%
{r q(y)+(1-r)p^{Y^{\rm\SSs id}}(y)}\!=:\!\frac{a_1 q(y)\!+\! a_2(y)}{a_3q(y)\!+\!a_4(y)}
\end{equation}
Hence we have to minimize
$$%\begin{equation}\nonumber
F(q):= \int \frac{|a_1 q(y)+ a_2(y)|^2}{a_3q(y)+a_4(y)}\,\,\mu(dy)
$$%\end{equation}
 in $M_0=\{q\in L_1(\mu)\,|\; q\geq 0,\; \int q\,d\mu=1\}$.
To this end, we note that $F$ is convex on the non-void, convex cone $M=\{q  \in L_1(\mu)\,|\; q\geq 0\}$ %---in fact even on $M'=\{q  \in L_1(\mu)\,|\; q\geq -a_4(y)/a_3\}$---%
so, for some $\tilde \rho\ge 0$, we may consider the Lagrangian
$$%\begin{equation}
L_{ \tilde \rho}(q):=F(q) + \tilde \rho \int q\,d\mu
$$%\end{equation}
for some positive Lagrange multiplier $\tilde \rho$.
Pointwise minimization in $y$ of $L_{ \tilde \rho}(q)$ gives % without any restriction gives us the form
%\begin{equation}
%q^0_s(y)=\Tfrac{1-r}{r} ( \big|D(y)\big|\big/s\,-1)\,\, p^{Y}(y)\nonumber
%\hat
$$q_s(y)=%\big(q^0_s(y)\big)_{\SSs +}=
\Tfrac{1-r}{r} ( \big|D(y)\big|\big/s\,-1)_{\SSs +}\,\, p^{Y}(y)%\nonumber
$$%\end{equation}
for some constant $s=s( \tilde \rho)=(\,|\Ew Y^{\ssr id}|^2 + \tilde \rho/r)^{1/2}$,
%and the restriction to $M$ leads to
%\begin{equation}
%\hat q_s(y)=\big(q^0_s(y)\big)_{\SSs +}=\Tfrac{1-r}{r} ( \big|D(y)\big|\big/s\,-1)_{\SSs +}\,\, p^{Y}(y)\nonumber
%\end{equation}
Pointwise in $y$, $\hat q_s$ is antitone and continuous in $s\ge 0$ and $\lim_{s\to 0[\infty]}q_s(y)=\infty[0]$, hence by
monotone convergence,
$$%\begin{equation}
H(s)=\int \hat q_s(y) \,\mu(dy)%\nonumber
$$%\end{equation}
 too, is antitone and continuous and $\lim_{s\to 0[\infty]}H(s)=\infty[0]$. So by continuity, there is some $\rho \in (0,\infty)$ with $H(\rho)=1$.
On $M_0$, $\int q\,d\mu =1$, but $\hat q_\rho=q_{s=\rho}\in M_0$ and is optimal on $M\supset M_0$ hence it also minimizes $F$ on $M_0$.
In particular, we get representation \eqref{P0def} and note that, independently from the choice of $\mu$,
the least favorable $P_0^{Y^{\rm\SSs di}}$ is dominated according to $P_0^{Y^{\rm\SSs di}}\ll P^{Y^{\rm\SSs id}}$, i.e.;
non-dominated $P^{Y^{\rm\SSs di}}$ are even easier to deal with.\\

As next step we %return to the minmax problem, i.e.; $\min_f{} \max_{\partial{\cal U}}[\ldots]$ and
show that
\begin{equation} \label{minmax=maxmin}
\max\nolimits_{\partial{\cal U}}\min\nolimits_f{} [\ldots] = \min\nolimits_f{} \max\nolimits_{\partial{\cal U}}[\ldots]
\end{equation}
To this end we first verify \eqref{f0def} determining $f_0(y)$ as $f_0(y)=\Ew_{\SSs {\rm re};\hat P}[X|Y^{\rm\SSs re}=y]$.
Writing a sub/superscript
``${{\rm re;}\,P}$'' for evaluation under the situation generated by  $P=P^{Y^{\rm\SSs di}}$
and $\hat P$ for $P_0^{Y^{\rm\SSs di}}$, we obtain the the risk for general $P$ as
\begin{eqnarray}
{\rm MSE}_{\SSs{{\rm re;}\,P}}[f_0(Y^{\SSs {\rm re},\,P})]&=&
(1-r)\Ew_{\rm\SSs id}\big|Y^{\rm\SSs id}-f_0(Y^{\rm\SSs id})\big|^2+ r \tr\Cov Y^{\rm\SSs id}+\nonumber\\
&&\quad  +
r\,\Ew_P \min(|D(Y^{\SSs {\rm di;},q})|^2,\rho^2) \label{9.7}
\end{eqnarray}
This is maximal for any $P$ that is concentrated on the set $\big\{\,|D(Y^{\SSs {\rm di;},q})|>\rho\,\big\}$,
which is true for $\hat P$. Hence \eqref{minmax=maxmin} follows, as for any contaminating $P$
$$%\begin{equation}
{\rm MSE}_{\SSs{{\rm re;}\,P}}[f_0(Y^{\SSs{{\rm re;}\,P}}] \le {\rm MSE}_{\SSs{{\rm re;}\,\hat P}}[f_0(Y^{\SSs{{\rm re;}\,\hat P}})]%\nonumber
$$%\end{equation}

Finally, we pass over from $\partial{\cal U}$ to ${\cal U}$: Let $f_r$, $\hat P_r$
 denote the components of the saddle-point for $\partial {\cal U}(r)$, as well as $\rho(r)$ the corresponding Lagrange multiplier
and $w_r$ the corresponding weight, i.e., $w_r=w_r(y)=\min(1, {\rho(r)}\,/\,{|D(y)|})$.
 Let $R(f,P,r)$ be the MSE of procedure $f$ at the {\rm SO} model
$\partial {\cal U}(r)$ with contaminating $P^{Y^{\rm \SSs di}}=P$.
As can be seen from \eqref{P0def}, $\rho(r)$ is antitone in $r$; in particular, as $\hat P_r$ is concentrated
on $\{|D(Y)|\ge \rho(r)\}$ which for $r\leq s$ is a subset
of $ \{|D(Y)|\ge \rho(s)\}$, we obtain
$$
R(f_s,\hat P_s,s )=R(f_s,\hat P_r,s )\qquad\mbox{for}\;r\leq s
$$
Note that $R(f_s,P,0 )=R(f_s,Q,0 )$ for all $P,Q$---hence passage to $\tilde R(f_s,P,r )= R(f_s,P,r )-R(f_s,P,0 )$
is helpful---and that
\begin{equation} \label{varsplit}
\tr \Cov Y^{\rm\SSs id}= \Ew_{\rm\SSs id} \Big[\tr \Cov_{\rm\SSs id}[Y^{\rm\SSs id}|Y^{\rm\SSs id}]+ |D(Y^{\rm\SSs id})|^2 \Big]
\end{equation}
Abbreviate $\bar w_s(Y^{\rm\SSs id})=1-\big(1-w_s(Y^{\rm\SSs id})\big)^2\ge 0$ to see that
\begin{eqnarray*}
\hspace{-2em}&&\tilde R(f_s,P,r )=
r \Big\{\Ew_{\rm\SSs id}\Big[|D(Y^{\rm\SSs id})|^2 \bar w_s(Y^{\rm\SSs id})\Big] +
\Ew_{P}\min(|D(Y^{\rm\SSs id})|,\rho(s))^2\,\Big\} \leq\\
\hspace{-2em}&&\leq r \Big\{\Ew_{\rm\SSs id}\Big[|D(Y^{\rm\SSs id})|^2 \bar w_s(Y^{\rm\SSs id})\Big] + \rho(s)^2\,\Big\}=
\tilde R(f_s,\hat P_r,r )<
\tilde R(f_s,\hat P_s,s )
\end{eqnarray*}
Hence the saddle-point extends to ${\cal U}(r)$; in particular the maximal risk is never attained in
the interior ${\cal U}(r)\setminus \partial {\cal U}(r)$. \eqref{sadvalSO} follows by plugging in the results.

\hfill\qed
\end{small}
%% ------------------------------------------------------------------------
%-----------------------------------------------------------------------------
%\section*{References}
%-----------------------------------------------------------------------------
%\bibliographystyle{plain}
\bibliographystyle{plainnat}
%-----------------------------------------------------------------------------

%
%-------------------------------------------------------------------------------
\end{document}